\documentclass[aps,superscriptaddress,nofootinbib]{revtex4}
\usepackage{epsfig,rotating}
\input{epsf}

\begin{document}

\title{The warm inflation story}

\author{Arjun Berera}
\affiliation{ School of Physics and Astronomy, University of
Edinburgh, Edinburgh, EH9 3FD, United Kingdom}

\begin{abstract}

Warm inflation has normalized two ideas in cosmology,
that in the early universe the initial primordial density
perturbations generally could be of classical rather than quantum origin
and that during inflation,
particle production from interactions amongst quantum field, and 
its backreaction
effects, can occur concurrent with
inflationary expansion.
When we first introduced these ideas, both were met with resistance,
but today they are widely accepted as possibilities with
many models and applications based on them, which is an indication of the
widespread influence of warm inflation.  Open quantum
field theory, which has been utilized in studies
of warm inflation, is by now a relevant subject in cosmology,
in part due to this early work.
In this review I first discuss the basic warm inflation dynamics.
I then outline how to
compute warm inflation dynamics from first principles quantum
field theory (QFT) and in particular how a dissipative term arises.
Warm inflation models can have an inflaton mass
bigger than the Hubble scale and the inflaton field excursion
can remain sub-Planckian, thus overcoming the most
prohibitive problems of inflation model building.
I discuss the early period of my work in
developing warm inflation that helped me arrive at these important
features of its dynamics.
Inflationary cosmology today is immersed in hypothetical models,
which by now are acting as a diversion from reaching any endgame in
this field. I discuss better ways to approach
model selection and give necessary requirements for
a well constrained and predictive inflation model.
A few warm inflation models are pointed out that could be developed
to this extent.
I discuss how at this stage more progress would be made
in this subject by taking a broader view on the possible
early universe solutions that include not just inflation
but the diverse range of options.

\medskip

\medskip

\noindent
keywords: early universe cosmology, warm inflation, quantum field theory, 
model building
\end{abstract}


\maketitle

\medskip

\section{Introduction}
\label{intro}

Warm inflation was introduced 28 years ago. 
At that time the standard inflation scenario, hereafter called
cold inflation, was overwhelming accepted as the valid description of
the early phases of the universe, with much anticipation of
its confirmation from the planned cosmic microwave background (CMB)
experiments within the coming decades.
In that time warm inflation has
gone from being considered by many in cosmology as a distraction
to one of the most promising solutions. 
The idea stems from
an elementary observation.
The central theme of
inflationary dynamics has been the evolution
of a scalar field, which during inflation carries most of the energy
of the universe and which interacts with other fields.
On the one hand, in the standard inflation picture
the tacit assumption made is that these
interactions have no effect apart from modifying the scalar
field effective potential through quantum corrections.
On the other hand, in the warm inflation picture
interactions not only do that
but also lead
to fluctuation and dissipation effects.
In condensed matter systems, interactions certainly
lead in general to all three of these effects
(some examples in \cite{weiss}).
Moreover from a statistical mechanics perspective,
the scalar field would want to dissipate its energy
to other fields, and the system as a whole would try
to equally distribute the available energy.
Ultimately a thorough dynamical calculation is needed
to address the question.
 
In cosmology, there is one important way this scalar field dynamics differs
from condensed matter systems, which is that all processes for
the former occur in an expanding universe.
Expansion acts to
constantly alter the state of the cosmological
system.  For example due to expansion,
radiation energy in the universe is continually
being diluted.  Similarly, the configuration
of any cosmological scale process is being altered over time.
Thus if the quantum mechanical processes that lead to dissipation
operate at a time scale much slower than the expansion rate
of the universe, than these processes would be totally shut down
due to expansion, even if in a nonexpanding system, like
a condensed matter system, the same processes operate efficiently.
This is the important question that must be understood.
In the early years of inflation, there was a viewpoint that
inflation had to be in a supercooled phase, since
expansion would be too fast for any such
microphysical processes to occur that lead to dissipation.
However our work in warm inflation changed this point of view.
Today this possibility is accepted without much question, 
thus one indicator of the wide influence and success of warm inflation.

The other major influence warm inflation has had is in normalizing
the possibility of the
initial primordial fluctuations being classical not quantum.
Again due to our timescale analysis, we showed there is considerable
dynamical range in the early universe for multiparticle processes,
such as those leading to thermalization or other statistical states.
When Li-Zhi Fang and I first were working on this idea  \cite{Berera:1995wh}, 
neither of us had a full
scenario in mind.  We simply wanted to demonstrate that
the prevailing idea of the times that primordial fluctuations,
whether during inflation or otherwise, had to
be quantum could be questioned.
When I first discussed these ideas back in the mid-90s, researchers
were surprised. 
I think one reason the initial paper
by Fang and me on thermally induced primordial fluctuations generated interest
when it came out was it was very novel for its time, yet
we presented the idea in a way that made it look familiar.
On its own this paper would probably have been forgotten
had it not been that a few years afterwards
with Marcelo Gleiser and Rudnei Ramos, we showed
within a quantum field theory model that relevant microphysical timescales
were possible to allow for such classical 
fluctuations \cite{Berera:1998gx,Berera:1998px}.
Over the years since then, there has been plenty of scrutiny as to whether
a full warm inflation scenario is viable, but the idea that
primordial fluctuations could be classical rather than quantum
had permanently taken root.  This changed the prevailing thinking about
primordial fluctuations dating back well before inflation,
that they are of quantum origin.
By now various ideas about
classical primordial fluctuations have been suggested,
and one of the successes of warm inflation has been that
the plausibility of such ideas is routinely accepted.

In this review I will discuss the warm inflation scenario and
the history of its development.  In the next two Sections
\ref{dynamics} and \ref{fpdynamics},
I will discuss the basic scenario and how to realize this
picture from first principles quantum field theory (QFT).  
Then  in Section \ref{bg}, I will
turn to the background of the idea, 
discussing my own steps in
realizing and developing warm inflation.  
In Section \ref{denpert}, I will discuss
density perturbations in warm inflation and the crucial difference
between the weak and strong dissipative regimes of warm inflation.
In Section \ref{mbuild}, I will discuss some of the first principles models of
warm inflation that have been constructed.  In Section \ref{advan}, I
will then talk about the advantages the warm inflation scenario has over
cold inflation. 
Intrinsic features about its dynamics allows warm inflation
to occur with the inflaton mass bigger then
the Hubble scale $m_{\phi} > H$ and the inflaton
field excursion less then the Planck scale $\phi < m_p$.
In Section \ref{critq}, a critique is given about 
difficulties warm inflation had faced in the field
for several years arising more from attitude then due to any
scientific shortcoming.
The particle physics consequences from
warm versus cold inflation are very different.  This makes it
important for CMB data to be viewed with a broad perspective,
with any lingering attitudes now best set aside.
I discuss warm inflation within the wider context of early universe
cosmology and the general direction in which this field it is headed.
In Section \ref{modselect}, I suggest ways to better select
the optimum inflation models.
There is a proliferation of inflation models in the literature
and some of this is taking the focus away from addressing
the key quantum field theory based problems to model building.
I suggest it would be useful to recognize the degree of speculation
in any given inflation model as one means to
separate out relevant models.
In the final Discussion Section \ref{diss},
I point out the varied types of
ideas and directions of research related to warm inflation,
which I was unable to cover in detail.  I also give some
closing comments. 

\section{The basic dynamics}
\label{dynamics}

Inflation in the most general terms is
a phase in which the scale factor grows at an accelerating rate,
${\ddot a} > 0$.  To derive inflation, one utilizes the
cosmological Einstein equations, for example
the scale factor equation
\begin{equation}
\frac{\ddot a}{a} = - \frac{4 \pi G}{3}(\rho + 3p) \;.
\label{scalee}
\end{equation}
It can be seen from this equation that
obtaining an accelerating scale factor ${\ddot a} > 0$,
requires $p < - \rho/3$, so a substance with
negative pressure.
This means a universe where the dominant form
of matter produces a repulsive form of gravity.
Vacuum energy has an equation of state $p_v = -\rho_v$,
which when dominant leads to inflation.
If the vacuum energy is the only energy in the universe and it is
constant, it leads to an exponential scale factor growth,
with this deSitter space being the most
common behavior associated with inflation.
In simplest terms, a universe undergoing accelerated expansion
grows much bigger in the same amount of time as a universe
undergoing decelerated expansion.

As it turns out, the equation of state of
a scalar fields contain terms with
negative pressure.  
The energy and pressure densities of a scalar field are
\begin{eqnarray}
\rho & = & \frac{{\dot \Phi}^2}{2} + V(\Phi) + \frac{(\nabla \Phi)^2}{2a^2}
\nonumber \\
p & = & \frac{{\dot \Phi}^2}{2} - V(\Phi) - \frac{(\nabla \Phi)^2}{6a^2},
\label{rhop}
\end{eqnarray}
so that a potential energy, $ V(\Phi)$, dominated state of
a scalar field has a negative pressure.
The idea generally adopted for realizing
inflation from particle physics is
to get the potential
energy of some scalar field to dominate 
the energy density
of the universe
for some short period of time in the early universe,
thereby generating the requisite
amount of inflation needed to solve the cosmological puzzles.
After sufficient inflation has then occurred, somehow place the
universe back into a radiation dominated hot Big Bang regime.
The scalar field that performs the task in driving
inflation is called the inflaton.
The inflaton must perform two essential roles.  These are
to supply an appropriate energy density to be
conducive for inflation and the inflaton fluctuations should
have the appropriate features to seed primordial density fluctuations in
the universe.

In both the warm and cold inflation pictures, to realize
inflation the scalar inflaton field
must be potential energy dominated.
The difference is, cold inflation is synonymous with
supercooling of the universe during 
inflation \cite{Guth:1980zm,Linde:1982uu,Albrecht:1982wi,Linde:1983gd}
(a good early review is \cite{Brandenberger:1984cz}),
whereas in warm inflation the inflaton is not
assumed to be an isolated, noninteracting field during
the inflation period.  This means, rather than the universe
supercooling, it
instead maintains some radiation during
inflation, to the extent
to noticeably alter inflaton dynamics.
In particular, the dividing point between warm and
cold inflation is roughly at
$\rho_r^{1/4} \approx H$, where $\rho_r$ is the radiation energy density
present during inflation and $H$ is the Hubble parameter,
which during the potential energy dominated inflation phase is
$H^2 = 8 \pi V/(3m_p^2)$, where $m_p$ is the Planck mass. Here
the warm inflation regime is $\rho_r^{1/4} > H$
and the cold inflation regime is $\rho_r^{1/4} \stackrel{<}{\sim} H$.
These criteria are independent of
thermalization, but if such were to occur, one sees
that the warm inflation regime basically corresponds to 
when $g^{*1/4} T > H$, where $g^*$ are the number of particle
species in the universe.
The relevance of this separation into these two regimes is since
the typical inflaton mass during
inflation is $m_{\phi} \approx H$, so that
when $T \stackrel{>}{\sim} H$, 
thermal or thermally induced fluctuations
of the inflaton field
will become important.

The interaction of the inflaton with other fields in general
implies its effective
evolution equation has terms representing
dissipation or emission of energy going out of the inflaton system into
other particles. 
Once the conceptual realization is made
that there could be the significant
level of particle production during inflation discussed
in the above paragraph, the whole
host of nonequilibrium QFT dynamics opens up
to consider.  The nomenclature of warm inflation
is meant to capture the kinematic property of the presence of
radiation without specific choice of production mechanism,
with this broad meaning discussed in the early 
papers \cite{Berera:1996fm,Berera:1999ws}.
Mainly warm near thermal equilibrium inflation
has been examined, since quite evidently
this is a difficult problem and that is the simplest case
to first study, although the requirement the inflaton move slowly
suggests this
may still be the best regime.  That will be the main focus in
this review and for brevity we will refer to it simply as warm inflation.
Note that the coupled 
set of Green's function equations for a general
nonequilibrium treatment of warm inflation has been developed
in early work by us and others \cite{Berera:2001gs,Berera:2007qm,Lawrie:2002wm,Lawrie:2002zd,Lawrie:2004hs},
including for
expanding spacetime \cite{Berera:2004kc,Moss:2008lkw}, 
as well as suggestions for an alternative
oscillation particle production based 
mechanism of warm inflation \cite{Bartrum:2013fia}
and adjusted thermalization \cite{Bastero-Gil:2017yzb}.
For warm near thermal equilibrium inflation,
the dynamics can be expressed as a
simple phenomenological Langevin type equation
\begin{equation}
{\ddot \phi} + [3H + \Upsilon] {\dot \phi} 
- \frac{1}{a^2(t)} \nabla^2 \phi 
	+ \frac{\partial V}{\partial \phi} = \zeta.
\label{wieom}
\end{equation}
In this equation, $\zeta$ is a fluctuating random force and
$\Upsilon {\dot \phi}$ is
a dissipative term.
Both these are effective terms that arise
due to the interaction of the inflaton
with other fields.  A fluctuation-dissipation relation
in general will relate these two terms, with
details depending on the microscopic dynamics and the statistical
state of the system.

For warm inflation to occur, the potential energy $\rho_v$ must be
larger than both the
radiation energy density $\rho_r$ and the inflaton's kinetic energy.
A major difference to cold inflation is the evolution of
the energy densities.  In warm
inflation, vacuum energy is
continuously being dissipated at the rate 
${\dot \rho}_v = - \Upsilon {\dot \phi}^2$ and so causing the radiation
energy not to vanish.
The General Relativity (GR) cosmological
energy conservation equation,
\begin{equation}
{\dot \rho} = -3 H (\rho + p) \;,
\label{econs}
\end{equation}
for this system of
vacuum and radiation becomes
\begin{equation} 
{\dot \rho}_r = -4H \rho_r + \Upsilon {\dot \phi}^2.
\label{radwi}
\end{equation}
The first term on the right-hand-side is a sink term that is
depleting radiation energy, whereas
the second term is sourcing this energy.
These equations are meant to
demonstrate the basic idea, so the notation is kept simple,
but to be clear in the above the field $\phi$ is just the
background mode, whereas in Eq. (\ref{wieom}) is represents
both the background and fluctuating modes.
Noting from the sink term that the rate of depletion is 
proportional to the amount
of radiation present, in general it implies there
will be a nonzero approximate steady state
point for $\rho_r$ controlled by the source term.  This holds
when $\phi$, $H$, and $\Upsilon$ are slowly
varying, which is a good approximation during inflation,
As an example, if the source term
is just a constant, which is a good approximation
during the slow roll evolution of the inflaton, then
$\Upsilon {\dot \phi}^2 = const. \equiv c_0$.
In that case the solution to equation (\ref{radwi}) will be
$\rho_r \approx c_0/(4H) + (\rho_{r0} - c_0/(4H)) \exp(-4Ht)$.
The second term on the RHS of this solution decays away
any initial radiation, but at large time radiation does not
entirely vanish because of the first term on the RHS.
Thus at large time,
the radiation in the universe 
depends only on the rate at
which the source is producing it
and so becomes independent of
initial conditions.

As already noted,
the presence of radiation during inflation
is fully consistent with the equations of General Relativity, since
the single requirement by the scale factor equation to realize 
an inflationary scale
factor growth is that the vacuum energy density is the dominant
component of energy in the universe.
This means supercooling is only one special limiting regime
of this general case implied by these equations.
Thus inflation would still happen if there was
say a 10\% or 1\% etc... admixture of radiation, in addition
to the vacuum energy.
This is an important point.  To appreciate it, note that there are
at least five
scales in inflation,
the vacuum energy $E_v \equiv \rho_v^{1/4}$,
the radiation energy $E_r \equiv \rho_r^{1/4}$,
the Hubble scale $H$,
the inflaton mass $m^2_{\phi} \equiv V''(\phi)$, and
the dissipative coefficient $\Upsilon$.
In the cold inflation picture, these five
energy scales are related as (i). $E_v \gg E_r$,
(ii). $H > m_{\phi}$,
(iii). $m_{\phi} > E_r$, and
(iv). $H \gg \Upsilon$.
Condition (i) is simply a minimal General Relativity requirement
to have inflation.
Condition (ii) is necessary for the slow roll regime.
Condition (iii) implies the universe is in a low-temperature
regime where radiation has an insignificant effect on inflaton fluctuations.
Finally condition (iv) implies
dissipation has an insignficant effect on
inflaton evolution.

For warm inflation, there are two regimes that
must be addressed, weak and strong dissipative warm inflation.
In both these regimes the following  energy scales are the same
(i). $E_v > E_r$,
(ii). ${\rm max} \ (\Upsilon,H) > m_{\phi}$, and
(iii). $E_r > m_{\phi}$.
Condition (i) is required again by General Relativity to
realize inflation.  Condition (ii) is the warm inflation
equivalent to the slow roll regime.
Condition (iii) implies the inflaton fluctuations are
no longer in a zero temperature state, so that radiation
will have nontrivial effects on inflaton dynamics and fluctuations.
Finally the last condition, and the one that leads to two
regimes of warm inflation, is
(iv). $\Upsilon > 3H$, strong dissipative warm inflation
and (iv). $\Upsilon \leq  3H$, weak dissipative warm inflation.
The notation here is almost self-explanatory, with the
strong dissipative regime, where the dissipative coefficient
$\Upsilon$ controls the damped evolution of the inflaton field and
the weak dissipative regime, where the Hubble damping still is
the dominant term.

Even if the presence of radiation does not hinder
inflationary growth, it can still influence inflaton dynamics.
For example
consider inflation at the
Grand Unified Theory (GUT) scale, so
$V^{1/4} \equiv E_v \sim 10^{15} {\rm GeV}$,
which means the Hubble parameter is
$H \sim V^{1/2}/m_p \sim 10^{11} {\rm GeV}$.
For cold inflation and weak dissipative warm inflation,
since the Hubble damping term $3H {\dot \phi}$ must
be adequate to produce slow roll inflaton evolution,
it requires that the inflaton mass
$m_{\phi} \sim 10^{9-10} {\rm GeV} \stackrel{<}{\sim} 3H$.
The key point to appreciate is that there are five orders
of magnitude difference here between the vacuum energy scale and
the scale of the inflaton mass.  In other words
there is a huge difference in scales between the
energy scale $m_{\phi}$ governing inflaton dynamics
and the energy scale $E_v$ driving inflation.
This implies, for example, 
that in order to excite the inflaton fluctuations
above their ground state, it only requires a minuscule fraction
of vacuum energy dissipated at a level as low as $0.001\%$.
This is good indication that dissipative effects during
inflation can play a noticeable role.
This is only an energetic assessment, but it is suggestive
of interesting physics.  It leaves then a question for a
full dynamical calculation to answer.  In particular,
the universe is expanding rapidly during inflation at a rate
characterized by the Hubble parameter $H$.  The question then is
whether the fundamental dynamics responsible for
dissipation occurs at a rate faster than Hubble expansion.

The other difference between warm and cold inflation is how dissipation
affects the parameters of the underlying first principles model,
which becomes most evident in the strong dissipative regime,
$\Upsilon > 3H$.  To understand this point,
recall that in cold inflation the inflaton motion is damped by
only the $3H {\dot \phi}$ term.  Thus slow roll
evolution requires the inflaton mass, $\sim \sqrt{ V''}$, to
be less than $\sim H$.  However in typical quantum field theory
models of inflation, it is very difficult to maintain such a
tiny inflaton mass, a point which is further addressed
in Sect. \ref{mbuild}.  In one form this is called
the ``$\eta$-problem'' \cite{Copeland:1994vg,Arkani-Hamed:2003mz}.
To contrast,
slow roll motion in warm inflation
only requires $V'' < (3H + \Upsilon)^2$,
so for $\Upsilon > 3H$ it means
the inflaton mass can be bigger
than in the cold inflation case, and in particular bigger than the Hubble scale.
This relaxation of the inflaton mass constraint permits
much greater freedom in building realistic inflaton models,
since this ``$\eta$-problem'', infra-red, and/or swampland problem 
is comfortably eliminated.

Another model building feature that differs warm versus
cold inflation is where inflation
occurs in regards the region of the scalar field background mode
amplitude, which is the zero-mode of the field
$\phi \equiv \langle \Phi \rangle$.
For cold inflation with the simplest
types of potentials, which also are the most commonly used,
$V = \lambda \Phi^4/4!$ and $V = m_{\phi}^2 \Phi^2/2$,
calculations show that the
initial inflaton amplitude has to be above the quantum gravity Planck scale
$\phi_i > m_p$.
This is because in these models, $H$ in the Hubble damping term,
$3H {\dot \phi}$, increases with larger field amplitude,
so in order to achieve an adequately long slow roll period
to yield the desired
50 or so efolds of inflation, 
this large field amplitude is required. However from the perspective
of the ultimate goal
of building a realistic particle physics
inflation model, this condition poses
a problem, which forces more complications into the
model building.  This will be discussed later in this review.
On the other hand in warm inflation, both dissipation and
radiation work to reduce the inflaton field amplitude.
Since thermal fluctuations will always be larger than quantum,
in order to constrain the scalar perturbation amplitude to be
the desired $\sim 10^{-5}$, it requires fixing the other parameters
such as the inflaton coupling and field amplitude
to be smaller, which in turn lowers the tensor to scalar ratio.
Moreover when $\Upsilon > 3H$,
this larger dissipation implies the 
the inflaton traverses a much smaller region of
the field amplitude in the slow roll phase, so
allowing its field amplitude to be smaller.
The basic point detailed calculations show 
is that for these simple monomial potentials,
in warm inflation the inflaton field amplitude can be
below the Planck scale $\phi < m_p$, thus avoiding
the quantum gravity scale.

\section{First principles dynamics}
\label{fpdynamics}

As already pointed out in the previous Section, 
the conversion of even a little
vacuum energy into radiation can have significant
effect during inflation.  Also Eq. (\ref{wieom})
has been presented as an effective evolution equation for the
inflaton field once interactions with other fields
are integrated out.  The questions that remains to be
answered from quantum field theory are whether both these
effects actually occur and if so then in what models.
In order to address these questions, in this Section our task
is to understand how to derive the effective equation
of motion for the scalar inflaton field 
starting from a fundamental Lagrangian. 

The basic Lagrangian quite generally for any inflaton model has the form 
${\cal L} = {\cal L}_S + {\cal L}_R + {\cal L}_I$. Here ${\cal L}_S$ 
is the inflaton system Lagrangian,
which has the general form
\begin{eqnarray}
{\cal L}_{S} =
\frac{1}{2} {\dot \Phi}^2 -
\frac{1}{2} (\nabla \Phi)^2 
- V(\Phi) .
\label{Lphipsi}
\end{eqnarray}
The inflaton in any model must interact with other fields,
since channels must exist from which 
the vacuum energy contained in the inflaton
field ultimately can be released into radiation energy, so that
inflation ends and the universe is put into a Hot Big Bang
evolution.  These interactions are contained in the ${\cal L}_I$ part of
the above Lagrangian.
The question is whether this conversion process
occurs exclusively at the end of inflation, as pictured
in cold inflation, or does it occur concurrent with
inflation, as pictured in warm inflation.
Some common types of interactions are the inflaton
coupled to bosonic fields such as $0.5 g^2 \Phi^2 \chi^2$
or fermion fields as $h \Phi {\bar \psi}{\psi}$,
i.e. $- {\cal L}_I = 0.5 g^2 \Phi^2 \chi^2 + h \Phi {\bar \psi} \psi$.
Finally ${\cal L}_R$ contains all other terms associated with all fields aside
from the inflaton that form the radiation bath or reservoir, like the
$\chi$ and $\psi$ fields in this example.

For this Lagrangian, the quantum operator equations of
motion can be immediately written down, one for
each field.   These equations are generally coupled to each
other due to nonlinear interactions.  We are interested in the
evolution equation of the fields, and in particular
the expectation value of the evolution equation
of the inflaton field,
given the state of the system at some initial
time $t_i$. 
Thus we wish to obtain the effective equation of motion
for the scalar inflaton field configuration 
$\phi \equiv \langle \Phi \rangle$, after
integrating out the quantum fluctuations in $\Phi$, and the effects of
all other fields with which $\phi$ interacts, like the
$\chi$ and $\psi$ fields in ${\cal L}_I$.
This is a typical ''system-reservoir''
decomposition of the problem, as familiar in
statistical mechanics \cite{weiss}.  In our
case the system is $\phi$
and the reservoir is all the other dynamical degrees of freedom.

The system-reservoir
approach has applications to many problems in physics.
It is instructive to state a few examples here. 
One of the most common examples is Brownian motion,
where the evolution of one singled out
particle is of interest, when it is immersed in a fluid
and interacts with particles in that fluid.  One seeks the 
evolution equation for this Brownian particle,
once the effect of all the other particles are integrated
out and represented in this equation through effective terms.
In our problem
the background field $\phi$ is the analog of the 
Brownian particle and
the reservoir bath in this case contains the $\Phi$ quantum modes, 
the scalar $\chi$, and the spinor
$\psi$.
In condensed matter
physics, the system-reservoir, or open quantum system, 
approach is widely used. Some examples are
the tunnelling of a trapped flux in a SQUID,
interaction in a metal of electrons with polarons, and
in Josephson junction arrays \cite{weiss,cmsr}.

In order to obtain the $\phi$ effective
equation of motion, the procedure is first to replace 
the field $\Phi$ in the Lagrangian by
$\Phi = \phi + \kappa$, where $\langle \Phi \rangle \equiv \phi$
and $\kappa$ are the quantum fluctuations
of the $\Phi$ field.  
Taking for example the potential 
$V= m_{\phi}^2 \Phi^2/2$,
the equation of motion for
$\phi$, then becomes
\begin{eqnarray}
\ddot{\phi} + 3H{\dot \phi}
+m_{\phi}^2 \phi - \frac{1}{a^2(t)} \nabla^2 \phi
+ g^2 \phi \langle \chi^2 \rangle + g^2 \langle \kappa \chi^2 \rangle
+ h \langle {\bar \psi} \psi \rangle = 0 \;.
\label{phiave}
\end{eqnarray}
One now wishes to solve the quantum operator equations of motion
for all the other fields, i.e.
$\kappa$, $\chi$ and $\psi$, as a function of $\phi$,
substitute these above in Eq. (\ref{phiave}), and
then take the specified expectation values.
What would emerge from this is the sought after effective evolution
equation for $\phi$.  This is in principle, but in practice
it can not be done exactly, so various perturbative and resummation methods 
are used.  In this review we will not
explore these approximation methods,
but the interested reader can examine
\cite{Berera:1998gx,Berera:2001gs,Berera:2007qm,Lawrie:2002wm,Lawrie:2002zd,Lawrie:2004hs,Berera:2004kc,Moss:2008lkw,Berera:2008ar}. 
Here only a few general features
of the effective $\phi$ equation of motion are highlighted.
First, since $\phi$ is singled out, it becomes an open system,
so it is expected that the $\phi$ effective 
equation of motion will be nonconservative.
Second, the fields $\chi$, $\psi$ etc...  at a given time
$t_0$ in general will be functions of $\phi$ at
all earlier times $t < t_0$.  Thus the expectation
values $\langle \chi^2 \rangle$ etc... in
Eq. (\ref{phiave}) will be nonlocal in time
with respect to $\phi$, so consistent with
the first general fact, as this will lead to a
nonconservative equation.  These time nonlocal terms are
then expressed in a derivative expansion of $\phi$ with respect to
time, and for adequately slow evolution, only
the leading term is retained to give the 
dissipative $\Upsilon {\dot \phi}$ term in Eq. (\ref{wieom}).

For the scalar field background mode, the emerging evolution equation
is simply ${\ddot \phi} + [3H+\Upsilon]{\dot \phi} + dV/d\phi = 0$,
which after multiplying through by a factor
of ${\dot \phi}$, can also be written in terms of the scalar field
Hamilitonian as 
$dH_S/dt = -[3H + \Upsilon] {\dot \phi}^2$ \cite{Berera:1996nv}.
This equation conveys that the loss in energy in the scalar inflaton
field sector is from the two terms on the right hand side,
one due to cosmological expansion and the other due to dissipation,
with the dissipative term then sourcing that energy to radiation,
as shown in Eq. (\ref{radwi}).
The  derivative expansion mentioned above means the leading
time nonlocal term associated with dissipation goes as
${\dot \phi}$, with $\Upsilon$, which has dimensions of rate
(energy dimension one), controlling
how fast the kinetic motion of the scalar field decays
its energy into particles due to its coupling to
other fields.  The QFT
formalism for computing $\Upsilon$ 
from the microphysical dynamics that folds into
producing this macroscopic dissipation term can be found in
\cite{Berera:1998gx,Berera:2007qm,Berera:2008ar,Bastero-Gil:2010dgy}.

For readers familiar with the effective potential in
Lagrangian quantum field theory, there is a heuristic way to understand
the origin of the $\phi$ effective equation of motion.
The effective potential calculation  
applies when $\phi$ 
is a static background.
The interaction of $\phi$
with the other quantum fields leads to
the creation of
quantum fluctuations, which are emitted off $\phi$, propagate
in space and time, and then are reabsorbed by $\phi$.
These processes typically are known as loop corrections, 
which modify the classical
potential and lead to the effective potential.
Now suppose $\phi$ is not in a static situation,
that it is changing in time.  In this case the same 
loop corrections mentioned above would occur. However at the time
of emission and absorption, the state of $\phi$
has changed.  Thus these loops no longer simply modify
the potential of $\phi$, but also introduce terms
which mix products of $\phi$ at different times,
therefore introducing temporally nonlocal terms into
the $\phi$ evolution equation.

Thus the key question is, given a particle interaction structure
in the Lagrangian, what types of
dissipative effects does this lead to during inflation.
In Sect. \ref{mbuild} we will review various models.
Just as one example, here is 
a two stage mechanism,
involving the inflaton field coupled to a heavy scalar field
$\chi$ which in turn is coupled to light fermion
fields $\psi$ as 
$g^2 \Phi^2 \chi^2 + h \chi  {\bar \psi}_{\chi} \psi_{\chi}$ 
\cite{Berera:2002sp,Berera:2003kg}.
In this case the background inflaton field $\phi$ acts
as a time dependent mass to the $\chi$ field. As $\phi$
changes over time, the $\chi$ mass changes, thus altering the
$\chi$ vacuum.  This leads to virtual $\chi$ production, which
then decay into real $\psi_{\chi}$ particles.  There are also direct
interactions of $\phi$ and $\chi$ particles.
This type of interaction structure is very common in 
particle physics models, thus conducive to warm inflationary dynamics.

\section{Background}
\label{bg}

My own background
in physics was not in cosmology. My PhD was on string theory and during
that time I also did considerable work in statistical physics
and condensed matter physics.  After finishing my PhD and during
my first postdoc in Tucson, Arizona, I started working
on perturbative QCD.  During that time I was the seminar organizer.
One academic who was in our department there was Fang.
I frequently had conversations with him and found
he had broad interests, with thought given to
many things.  So I asked him to give us a seminar and he said
he would on inflationary cosmology.  This was not a subject I had studied
before.  During his talk as he discussed the scalar inflaton field
and how it can drive inflation, I raised my hand and asked him
where the dissipative term was in the inflaton evolution equation.
Coming from a background that included condensed matter physics,
I found it unusual that an interacting system did not have a
term accounting for dissipation. Fang paused his talk and said
I should come by afterwards for a discussion.  So I did and I
explained to him how I'd expect the inflaton evolution equation
to have a standard dissipative term and possibly even be
governed by a Langevin type evolution. This seemed to be a direction
Fang had given thought to before. He showed me a paper
he had written in 1980 which had basically 
suggested inflation \cite{Fang:1980wi}.  It also had
radiation production during inflation, but neither of us
felt the dynamics of that paper was the direction to develop further.

The basic idea of an exponential expansion phase in the early
Universe was first suggested in the highly insightful work
by Gliner starting in the mid-1960s \cite{Gliner1,Gliner2}.  
Then during the 70s,
Kirzhnits and Linde developed the foundations for application of
particle physics to phase transitions in cosmology,
including explaining how they could be important in explaining
the cosmological puzzles 
\cite{Kirzhnits:1972iw,Kirzhnits:1972ut,Kirzhnits:1974as,Kirzhnits:1976ts,Linde:1978px}.  
By the late 70s and early 80s there
were several papers using these seminal ideas including Fang's paper.
His was one of the set of papers prior to Guth's \cite{Guth:1980zm} that
had suggested the inflation idea as a solution to
the horizon problem 
\cite{Zeldovich:1968ehl,Brout:1977ix,Starobinsky:1980te,Kazanas:1980tx,Kolb:1979bt,Sher:1980qu,Sato:1980yn}, 
but without the catchy name that Guth finally
gave the scenario (Guth's paper also pointed out that inflation could
solve the flatness problem).
All this work aside from Fang's was developing cold inflation
dynamics.

Fang and I started
developing our ideas and wrote the paper \cite{Berera:1995wh}.
This addressed dissipation and noise and came up with an expression for 
inflaton fluctuations that were thermally induced.  At that point my postdoc
there was coming to an end, and I was moving to Penn State.  During
that period I started thinking about formulating a full scenario
and showing one could realize inflation and have dissipation
concurrently with these thermally induced fluctuations.  I did eventually
arrive at a model and I sent my results to Fang.  I had naturally
considered him a collaborator on this work. Fang found the results
very interesting and encouraged me to write it up and publish
it \cite{Berera:1995ie}.
However he said he had not thought about our idea to this extent
and so I should write this paper on my own.
He and I even had a discussion about
what to name this new scenario, which led to 
calling it warm inflation.

At the time I considered this work as just a side project to my
main interests developing in perturbative QCD. However I kept thinking
about it and wrote a couple more papers in the next couple of
years.  In one paper I studied warm inflation trajectories computed from
the Friedmann equations for a system with both a decaying vacuum energy and
radiation, and showing how such evolution could smoothly go from
a warm inflationary phase to the radiation dominated regime, thus
offering a graceful exit \cite{Berera:1996fm}.  
I also started getting interested in
how an inflaton Langevin type equation could be derived
from first principles. After all my initial comment in Fang's
talk had been that there should be a dissipation and noise term
in such equations.  In my first attempt to understand 
such dynamics \cite{Berera:1996nv},
resting on my condensed matter background, I studied
the Caldeira-Leggett model \cite{cmsr}.  
The original was a quantum mechanical
model of a single coordinate, the system, coupled linearly to many
other coordinates, all being harmonic oscillator Hamiltonians.
My paper made a quantum field theory extension of this model to study
inflationary expansion concurrent with dissipation, which
set some foundations for a QFT derivation of warm inflation dynamics.

Around this point I was looking for a new postdoc position.  By now my
research interests had turned heavily toward developing warm inflation
and not many people were interested in the idea, so I had difficulty
getting hired.  However Robert Brandenberger responded
to my efforts by reaching out in an email to me saying he
found my ideas interesting. He was one of
the first researchers in cosmology
to support my work.  I am not sure that Robert believed warm
inflation was necessarily THE idea of 
cosmology. I think his attitude, like mine,
is that any reasonable idea in cosmology needs to be fully
examined.  There is no way that any
idea in cosmology can become the single adopted picture
until all reasonable ideas have been fully vetted.
In that context I think he felt warm inflation
deserves its time to be developed and considered.
Through his help I was able to secure a postdoc
position at Vanderbilt in the group of Tom Kephart and Tom Weiler.
It was here that my work in warm inflation developed considerably.
They had a very open minded attitude toward developing new
ideas in theoretical physics and were not too bothered about
just following the mainstream.  This provided a conducive
environment.

After writing these initial papers
on warm inflation,
I was contacted by Gleiser and Ramos.
They had written one of the pivotal papers in 
deriving dissipation in a scalar
quantum field theory \cite{Gleiser:1993ea}
(other papers on scalar field dissipation
around or previous to this are 
\cite{Hosoya:1983ke,Morikawa:1984dz,Morikawa:1987ci,Ringwald:1987ui,Calzetta:1986cq,Boyanovsky:1994me}). 
They saw the connection between their work and
what I was trying to achieve. We started working together and
wrote a paper of a scalar field $\phi$, meant to be the inflaton, coupled
to $N$ other scalar fields which are integrated out to arrive
at an effective equation of motion 
for $\phi$ \cite{Berera:1998gx}.  This effective equation
of motion would contain then a dissipative term.  This model
contained all the basic features for realizing warm inflation
dynamics.  Because all masses were much bigger than the Hubble
scale and all microphysical dynamics was fast compared to
the expansion scale, this model could be calculated still within
a flat spacetime framework. This is a noteworthy feature of
warm inflation that on a relative scale for a nonequilibrium open
QFT problem,
the underlying dynamics is fairly simple
to calculate.  These conditions on the field theory were imposed as
consistency conditions. In particular we introduced adiabatic
conditions that the dynamic time scale of evolution of
the scalar field be much larger than typical collision
and decay time scales $\Gamma^{-1}$,
\begin{equation}
\phi/{\dot \phi} \gg \Gamma^{-1} \;.
\end{equation}
We also imposed that
these microphysical collision and decay time scales be much
shorter than the Hubble time, 
\begin{equation}
\Gamma \gg H \;. 
\end{equation}
As I mentioned earlier, a key initial barrier to why no one before me
had suggested the warm inflation scenario seems to have
been due to a lack of understanding
of dynamical time scales relevant to inflation and in particular
most researchers in this field held to a belief that inflation 
happens too quickly for particle production
to occur.  In this paper with Gleiser and Ramos
my initial thoughts about such time scales
was examined within quantum field theory, and we recognized that
indeed timescales can allow for particle production, although
it would not be easy to realize. This was
one major accomplishment of this paper.

Several months after we put this paper on the arxiv, 
Junichi Yokoyama and Andrei Linde (YL)
wrote a paper titled `Is warm inflation possible?' and seemed
to have answered their question within a one sentence abstract that
said it was `extremely difficult and perhaps 
even impossible' \cite{Yokoyama:1998ju}.
The analysis in their paper was not that different from ours.
They did compute the fermonic channel of dissipation whereas our
paper had done the bosonic, so that added a useful new result.
However their basic analysis of warm inflation followed
ours as did the consistency conditions. Only the conclusions
differed. Where they saw impossible we simply saw a set of constraints
that would help guide us towards building a first principles
model of warm inflation.

After our paper but before Yokoyama and Linde's, 
Ramos and I had presented our work at
PASCOS98 in Boston.  
Linde was in the audience for our talks and told us
that he did not think our warm inflation work was correct,
although subsequently still during the conference he told us
there was some merit to our dissipation calculations.
Nevertheless some months later he
wrote the above mentioned paper.  Yokoyama and Linde did
share the results of their upcoming paper with us
with this claim about the impossibility of warm inflation.
Thus, we got to work on building a quantum field theory model that
demonstrated warm inflation is possible.  Within a few
weeks after they arxived their paper, we put out the first
quantum field theory model, which we called the distributed
mass model (DMM, explained below in Sect. (\ref{mbuild})), 
demonstrating that the
`impossible' was really not quite so \cite{Berera:1998px}.  
The following year
Tom Kephart and I built a string theory motivated realization of
this model \cite{Berera:1999wt}, further solidifying that not only
is warm inflation possible but it has attractive
model building prospects.

Around this time my interest turned to taking a deeper examination of
warm inflation dynamics. The regime of warm inflation that to me
seemed most interesting, and still does, is the strong
dissipative regime of warm inflation
\begin{equation}
\Upsilon > 3H \;, 
\end{equation}
so that the
damping of the inflaton motion was dominated by the thermal
damping and Hubble damping did not play a major role.  In the first
paper with Fang, we had computed the density perturbations only
in the weak dissipative regime of warm inflation
\begin{equation}
\Upsilon \leq 3H \;.
\end{equation}
In \cite{Berera:1999ws} I determined the 
expression for the density perturbations
in the strong dissipative regime.  In this paper I also looked
more closely at the quantum field theory dynamics using
the distributed mass model we had introduced. One of
the key observations I made in that paper was that in the
strong dissipative regime $\Upsilon > 3H$,
the mass of the inflaton could be larger than the Hubble
scale.  This allowed for warm inflation models unlike anything
that could be made for cold inflation, where slow roll under
Hubble friction required that $m_{\phi} < 3H$.  A mass less than
$H$ implies a Compton wavelength bigger than the horizon.  For
such a case, in the rest frame the particle is not localizable,
so the field associated with it has no particle interpretation
within the conventional sense.
Thus a matter field with $m_{\phi} < H$
was unlike any kind of quantum field we had any empirical
knowledge about from collider experiments
(Note that photons being massless have no rest frame and so are 
not localizable,
but this is due to their Abelian gauge symmetry which makes them very different
from material particles like from a scalar field.).
It opened
up the possibility for infra-red problems.  So one of the conditions
I imposed for what I considered the ideal inflation
model was that $m_{\phi} > H$ 
and I called it the infra-red condition \cite{Berera:1999ws}.
I had also understood that dissipation could lower the
background inflaton field amplitude and in particular
for monomial potentials could allow $\phi < m_p$ \cite{Berera:2003yyp}.
Both these conditions I recognized by the early 2000s
from simple reasoning as being important
for an ideal inflation model, well before they emerged in the swampland
conditions \cite{Obied:2018sgi,Ooguri:2018wrx}.
These conditions set the goal for what to look for in deriving 
a warm inflation model from
first principles quantum field theory.  There has been some
success in this direction, which I will discuss in Sect. (\ref{mbuild}).
However it has proven very difficult to find
models in this regime. Nevertheless in principle it is possible,
which is very different from cold inflation where such a regime
is very difficult to achieve and in particular having an inflaton
mass larger than the Hubble scale is out of the question.
It remains an open model
building question for warm inflation to find such models.

At this point Rudnei and I set out to build more first principles
warm inflation models.  Initially we explored supersymmetry (SUSY), which
was one of the most well utilized symmetries in inflation model
building to realize the ultraflat inflaton potential that
was needed. In warm inflation there were the dual requirements for
having this very flat inflaton potential but at the same time
creating a large enough dissipative term.
This effort resulted in  the SUSY two-stage 
dissipation model \cite{Berera:2003kg} (further explained in
Sect. (\ref{mbuild})).
We computed the dissipative coefficient, the radiative
corrections, and consistency conditions for this model and obtained
a warm inflation regime, although like DMM once again this model
required a large number of fields going upward of thousands.
At around this time, I also started working with Mar Bastero-Gil
and we further examined this and other 
SUSY models \cite{Bastero-Gil:2006ahd}. 
Also around this time Ian Moss got interested in warm inflation, 
with Hall, Moss, and Berera \cite{Hall:2003zp} doing 
a more detailed examination of density
perturbations.  He continued to study this even
further, with Graham and Moss \cite{Graham:2009bf}
finding a certain growing mode for the fluctuations if
the dissipative coefficient was temperature
dependent.
This result added yet more numerical difficulties in computing
warm inflation from a model.  Bastero-Gil and Ramos developed a set
of codes that could do the needed calculations.
The warm inflation power spectrum would have contribution from
both quantum and thermal noise. Ramos and da Silva \cite{Ramos:2013nsa}
did a careful analysis of these contributions starting
with my basic expressions for the primordial
density perturbations and came up
with a total power spectrum.

Alongside these developments of the theory, we did model building and
made predictions for the CMB from the two-stage mechanism model.
In 2009 the paper by Bastero-Gil and me \cite{Bastero-Gil:2009sdq} 
demonstrated that in warm inflation
the tensor-to-scalar ratio, $r$, would be suppressed compared to
the comparable cold inflation model for the monomial
potentials $\Phi^2$ and $\Phi^4$.  At the time this was a result that
went against the growing tide of expectation of finding a high
scale tensor mode at the Grand Unified Theory scale.  In 2014 our 
paper \cite{Bartrum:2013fia} did
a more detailed
study to show the suppression of the tensor mode for
the warm $\Phi^4$ model and also consistency for $n_s$ with the
Planck 2013 results.  This paper also showed that as the dissipative
coefficient increased, $r$ would decrease, thus 
demonstrating again the parametric
suppression of the tensor mode with increasing dissipation.
This was a significant finding. Although by this time there was a
trend in CMB data that the upper bound on $r$ was decreasing, it was edging on
ruling out the cold inflation $\Phi^4$ model, there was anticipation
that a tensor mode would be found. Our results went contrary to such
expectations and indicated that the tensor mode would be
suppressed.

Although the two-stage model was very successful in developing
a working warm inflation model, the fact it required a huge number
of fields was an issue we were not too happy about. We wanted to find
a simple warm inflation model, that contained only a small
number of fields.  Our efforts in doing that with SUSY we felt had
been exhausted, so we started exploring other symmetries.
One that we had been thinking about for some time was the pseudoscalar
symmetry.  This led us to construct the warm little 
inflaton model \cite{Bastero-Gil:2016qru}. This model will
be discussed in more detail below in Sec. \ref{mbuild}, but in short
it obtained warm inflation with just a few fields, thus was a
major step forward in constructing successful particle physics models
of warm inflation.

There had been work previous to my warm inflation paper in 1995 and that
by Fang and me earlier that year, 
which had discussed dissipation during
inflation. To start with was
Fang's 1980 paper \cite{Fang:1980wi} that I already mentioned.
He examined a source of dissipation associated with bulk viscosity
specific to a phase transition and was not the direction
that seemed could be developed in any detail.  Subsequently
in the mid-1980s,
Moss \cite{Moss:1985wn} and Yokoyama 
and Maeda \cite{Yokoyama:1987an} suggested the idea
of dissipation in the inflaton evolution equation similar to
warm inflation, though we were not aware of these two
papers when initially developing warm inflation. 
The success of warm inflation gave
these interesting early works 
a new lease on life.
However none of this early work appreciated the importance of
time scales, so that the dissipation could only be present if
the microphysical dynamics producing it operated faster than the
macroscopic time scale of expansion. 
This is really the key question to answer as to whether
warm inflation is a viable idea.
Nor did these early works pick
up on the underlying fluctuation-dissipation dynamics of warm inflation,
and that property is more general than just the thermal limit.
Finally an expression for density perturbation was obtained in Moss's paper,
but it was only for the weak dissipative regime, with his paper not
understanding the distinction between dissipative regimes.
In fact Fang and I made a similar oversight in
deriving the same expression and thinking it was
generally valid, when actually it was
only for the weak
regime \cite{Berera:1995wh} unaware of
the Moss paper \cite{Moss:1985wn}.
Yokoyama and Linde \cite{Yokoyama:1998ju} had briefly 
commented that
the presence of the dissipative coefficient in the inflaton
evolution equation may affect the expression for the density
perturbation but gave no details and just used the expression
of Fang and me.
It was only after a few years of studying warm inflation
did I realize there are two very distinct regimes of warm inflation,
strong $\Upsilon > 3H$ and weak $\Upsilon \leq 3H$.
I then understood that Fang's and my
original expression for the density perturbation was only
valid in the weak regime, and I then obtained
the expression for the density perturbation
in the strong regime \cite{Berera:1999ws}.
It is the strong regime that has the most interesting features of warm
inflation, since with $\Upsilon > 3H$ it allows $m_{\phi} > H$,
thus cleanly solving the $\eta$-problem \cite{Berera:2003yyp}.
Also the strong regime can allow $\phi < m_p$, thus allowing
all scales in the model to be below the quantum gravity scale.
In more recent terms this is the best regime to overcome
\cite{Das:2018rpg,Motaharfar:2018zyb,Berera:2019zdd,Berera:2020dvn}
all the swampland 
difficulties  \cite{Obied:2018sgi,Ooguri:2018wrx,Bedroya:2019tba}.

\section{Density perturbations}
\label{denpert}

In the most common realization of warm inflation,
density perturbations are induced from a thermal
bath.  They are classical on creation and thus the scenario has no
quantum-to-classical transition problem as is the case for cold inflation.
In cold inflation the inflaton density perturbations are dictated by
the Hubble scale where modes freeze-out to 
give $ \delta \phi \sim H$.  In contrast
in warm inflation there are three scales, the Hubble scale,
the dissipation scale, and the
temperature during inflation.
The dissipative term, $\Upsilon$, in warm 
inflation can be much larger than the Hubble damping 
term $H$ during
inflation. Due to the $\Upsilon$ term, this 
freeze-out momentum scale can be
much larger than that in cold inflation, which is $\sim H$.  At the freeze-out
time $t_F$, when the physical wavenumber $k_F=k/a(t_F)$, the mode amplitude
$\overline{\delta\phi}$ can be estimated using a purely thermal spectrum,
\begin{equation}
\overline{\delta   \phi}^2(k_F)  \approx   \int_{k<k_F}  \frac{d^3k}{(2\pi)^3}
         {1\over\omega_k}(e^{\beta\omega_k}-1)^{-1}   \stackrel{T  \rightarrow
           \infty}{\approx} \frac{k_F T}{2 \pi^2} \;.
\label{dphiest}
\end{equation}
To estimate $k_F$, one must determine when the damping rate of Eq.
(\ref{wieom}) falls below the expansion rate $H$, which occurs at
$k_F^2\approx (3H+\Upsilon)H$.  Thus, in the strong dissipative regime 
$Q \equiv \Upsilon/(3H) \gg 1$,
this implies $k_F \sim \sqrt{H \Upsilon}$.  Substituting for $k_F$ in Eq.
(\ref{dphiest}), one finds the expression for 
the inflaton fluctuation amplitude at
freeze-out
\begin{equation}
\overline{\delta \phi}^2 \sim \frac{\sqrt{H \Upsilon} T}{2 \pi^2} \;.
\label{dphistrong}
\end{equation}
This expression was first derived by me in \cite{Berera:1999ws}.
In the weak
dissipative regime $Q \ll 1$, the freeze-out wavenumber $k_F \sim H$,
which is consistent with cold inflation, thus giving the inflaton
fluctuation amplitude at freeze-out,
\begin{equation}
\overline{\delta \phi}^2 \sim \frac{H T}{2 \pi^2} .
\label{dphiweak}
\end{equation}
This expression was found by Moss \cite{Moss:1985wn} 
and then independently
rediscovered by Berera and Fang \cite{Berera:1995wh}.
In both cases
the regime of its validity was wrongly understood, and in
\cite{Berera:1999ws} the appropriate regime in which it was valid, the weak
dissipative regime, was clarified.

The fact that the density perturbations in warm inflation are classical
and of thermal origin some regard as an unappealing picture for
the early universe, but there is no concrete argument behind
this attitude.  The idea that the initial
primordial perturbations are quantum in origin has over the years
become encased in the lore of early universe cosmology, but
the factual basis for having such a beginning is lacking.  
There was initially some ideas about the universe being created
as a quantum fluctuation.  Moreover the 
chaotic inflation model \cite{Linde:1983gd}
did provide some kind of
dynamical picture motivating the origin of quantum fluctuations.
However this model required inflation at a very high energy at
the GUT scale, which
has been ruled out by CMB Planck data \cite{Planck:2015fie}
and further constrained by
more recent BICEP data \cite{BICEP:2021xfz}.

There
has been furious effort over the decades to develop ever more
cold inflation models with their own unique signatures from the
density perturbations. However the data itself has been far from
revealing and the fact of the matter is what is seen
there can equally be explained by the quantum fluctuations of
cold inflation or the classical ones of warm inflation. 
So far observational
data shows absolutely no preference.  Moreover, of the myriad of
possible interesting effects that might emerge from density
perturbations, it is a model building game and one can concoct
various such features both from warm and cold inflation models.
More recently there has been work to look for intrinsic
features that signal a quantum origin or in its absence a classical
origin \cite{Maldacena:2015bha,Green:2020whw,Brahma:2021mng,Dale:2023fnp}. 
However such tests appear to be extremely difficult
and if the CMB data remains without significant features,
maybe impossible to decisively measure.
The real lesson that can be taken away from these papers
is just how hard it is to discriminate between quantum versus
classical primordial perturbations during inflation.  This makes
it all the more perplexing how some adhere 
to the belief that the primordial fluctuations must have
been quantum.
In actuality there is equal reason to believe 
both classical and quantum processes
play roles in phenomenon in the early
universe.  This is the correct unbiased initial assumption
that should be taken for a robust examination of early universe cosmology.

This adherence to the primordial fluctuations being quantum is
more a statement about present attitudes in theoretical physics.
This has historical parallels.
Somewhat more than a century back, the established thinking was
that the world was governed by deterministic classical physics.
By now attitudes in theoretical physics tend to almost the other extreme and
identifying quantum phenomenon where possible is all the rage.
We are more fortuitous compared to the state of physics a century back
in that we have an extensive understanding of both classical
and quantum physics.  
The rational attitude is
to accept both possibilities for the origin of density perturbations and
let the science decide.  The early universe was large enough
to allow for classical behavior. 
In fact for anything bigger than the quantum gravity scale,
there is no argument to favor quantum fluctuations over classical ones.
From what we know, inflation had to occur below the Planck scale,
below the string scale and even below the lower end of the GUT scale.
As the bounds on
the tensor mode decreases, if interpreted in terms of inflationary
dynamics, meaning the energy scale of inflation decreases, so
moves even further below the quantum gravity scale, the arguments for
a quantum origin of perturbations become less compelling.
The universe may still have initially emerged from some type of
quantum gravity scale fluctuation, but after that
the ensuing dynamics need
not all then be quantum. Its well possible that particle production
occurred and density fluctuations then had a classical characteristic to
them whether thermally induced, thermal, or any other statistical state.
There is also an intermediate possibility that these primordial
fluctuations have mixed quantum and classical properties.

The idea of quantum fluctuations seeding the initial density perturbations
was suggested early on in the 1950s by 
Wheeler \cite{Wheeler:1957mu} and later considered
by Harrison \cite{Harrison:1969fb}.
In these early works there was no
mechanism suggested for producing such fluctuations, but
it was simply asserted that
their presence could explain what was known at the time about
large scale structure.  A noteworthy point about 
these early papers is they assumed
these fluctuations would have been created at the quantum
gravity scale $m_p$, with their simple argument being that at that
scale classical physics fails.
Linde's chaotic inflation scenario
offered a mechanism that linked quantum gravity scale physics down
to the GUT scale, where he postulated quantum fluctuations.
His scenario did not require at that stage quantum fluctuations.
There were known dynamical QFT models at that scale, so some dynamical
process such a thermalization etc... could still be conceivable.
At the moment the most stringent bound on the tensor-to-scalar
ratio is from BICEP, placing
the current upper bound of $r\sim 0.03$ \cite{BICEP:2021xfz}. This would
correspond to an energy scale during inflation less than 
$\sim 10^{15} {\rm GeV}$. This implies a Hubble scale during inflation
$H \sim \sqrt{V}/m_p \approx 10^{11} {\rm GeV}$, which is
seven orders of magnitude above the Large Hadron Collider energy
scale. From our developed theoretical knowledge about the early universe,
this does not come across as a particularly
fast timescale.  Moreover at this conceivable energy
scale $\sim 10^{15} {\rm GeV}$
for inflation, there are plenty of particle physics
models that have been constructed, GUT etc..., that
could provide degrees of freedom operating fast enough
to create a thermal or some type of multiparticle statistical
state leading to classical fluctuations. Thus there is no
reason to expect at this conceivable energy scale of
inflation or lower, which is much below the quantum gravity scale,
that primordial fluctuations must be uniquely quantum.
Should a tensor mode eventually be found, based on the present
bounds, we know the corresponding energy scale will
be below the GUT scale. In such a case, the arguments
are very compelling that such a outcome is
favoring warm not cold inflation.

The gamble taken by cold inflationary cosmology was that 
the tensor mode signatures
for inflation would be found at the GUT scale, based on an
chaotic inflation explanation involving the
simple monomial potentials. 
Nevertheless, there always was ample grounds to be cautious
about these models, since their predictions came from a questionable
regime of QFT
with a sub-Hubble inflaton mass and super-Planckian field
excursion,
and eventually the data ruled them out.
With the trends in
the data not supporting the simple monomial cold
inflation models, it's best now to be more
open-minded.  Meaningful progress in these theoretical questions about
the early universe will only happen by taking a broad
view.  If data does eventually confirm a low tensor-to-scalar ratio,
there are strong arguments that it is confirming warm rather than cold
inflation.  

In order to
decide which is the more compelling origin of density perturbations
thus the more compelling inflation picture,
the data alone will not be sufficient, since there is only
a limited amount of information we can measure
about such an early time period of the Universe.  Equally
it needs to be seen which scenario is most compelling from
a theoretical perspective, a point that
will be discussed in greater detail in Section \ref{modselect}.  A minimal
requirement has to be that the scenario can be 
cleanly derived from
quantum field theory.  The success of quantum field theory
in collider physics implies this is the best and only tool
we have to explore the high energy regime. And for energy regimes yet
far beyond measurement, the best we can do is rely on
the rules of quantum field theory that we have learned at these
lower energy scales. And if some consistent picture based on
those rules emerges for higher energies, then that
is the best possible prediction we can make.  
This of course means building a model beyond the Standard Model (SM),
however using types of fields and if possible even symmetries that are
known in the Standard Model. In particular such a model
should not rely on gravity, since we know nothing definitive about
its quantum nature and even have limited knowledge about
its classical nature.
In this respect
the cold inflation scenario, though simple in appearance, hides
many problems. We have already mentioned the problems that
emerge due to the inflaton mass being less than the Hubble scale.
Whether they are infra-red, $\eta$, or swampland problems, this
small mass scale seem something unwanted by quantum field theory.
Likewise a scalar field amplitude above the Planck scale
introduces unknown quantum gravity concerns.  Then there are
quantum-to-classical transition issues.

One note here on terminology. Nowadays many researchers refer to the type
of quantum field theory the Standard Model is built on as
an effective field theory, suggesting it is subservient to some
higher theory. Nevertheless to date there is no such established higher
theory. This is all still a matter of research and speculation.
This type of nomenclature is fine as a matter of convenience
for those working on higher theories. However when talking about
predictions and comparing to experiment, it can be misleading.
It can suggest the quantum field theory we know and understand is
somehow less predictive than the higher theory.
Until there is an established higher theory, 
the quantum field theory we know,
and the rules it embodies, is
the most predictive tool we have. 
As such in this review
I will refer to the quantum field theory of the Standard Model
as simply quantum field theory, first principles quantum field theory,
conventional quantum field theory,
or the quantum field theory we understand etc...  I will include
in this terminology effective field theories that
have cutoff scales below the Planck scale, such as sigma models
and such models involving pseudo Nambu-Goldstone bosons
or other models built on symmetries found in the Standard Model.

\section{Model building}
\label{mbuild}

The ultimate goal of warm inflation model building is to find models
computed from first principles quantum field theory.  This requires
that the model produces dissipation and the microphysical dynamics
operates faster than the macroscopic dynamics, so faster than
the evolution of the inflaton field 
and the expansion rate, $H$, of the universe.
These requirements emerge as consistency conditions in a warm inflation
calculation. Finally once a working model has been developed, one
then needs to check whether its predictions are consistent
with observation.  Achieving all this is a very difficult task
and so far only a few such warm inflation models have been developed.

Alongside this first principles QFT model building, there also
has been phenomenological warm inflation model building.
In this approach one simply puts in by hand a dissipative coefficient
in the inflaton evolution and then computes the resulting warm inflation.
This approach is useful for exploring types of dissipative behavior
that can lead to observationally consistent warm inflation
models.  Given how hard the first principles approach is,
this approach provides an intermediate step to studying
the types of warm inflation models that could be relevant.

Here I will discuss some of the first principles quantum field
theory warm inflation models that have been developed.
The first such model is what we called the distributed 
mass model (DMM) \cite{Berera:1998px}.
In these models there are a set of bosonic fields $\chi_{i}$
which interact with the inflaton field through shifted couplings.
The interaction
term in the Lagrangian which realizes such shifted couplings has the form,
\begin{equation}
\frac{g^2}{2} (\Phi - M_i)^2 \chi_i^2 ,
\label{dmmchiint}
\end{equation}
so that when $\langle \Phi \rangle = \phi \sim M_i$, 
the $\chi_i$ field mass
becomes small.  In
particular when the mass of a $\chi_i$ field gets below 
the temperature scale in the
Universe, it becomes thermally excited.  Once
thermally excited, as the background inflaton field evolves, it is able to
dissipate energy into these fields.  This creates a dissipative term in the
inflaton evolution equation \cite{Berera:1998gx}.  
As an aside, the idea of these shifted 
couplings of the inflaton in our DMM
model has subsequently been used to develop other 
types of warm inflation models
including trapped 
inflation \cite{Kofman:2004yc,Green:2009ds,LopezNacir:2011kk}.

For the DMM, if these mass scales $M_i$ are now
distributed over some range that $\phi$ will traverse, then during
evolution of $\phi$, some subset of these $\chi$ fields will be light and
generate a dissipative term.  In order to control the radiative corrections,
this needs to be extended to a Supersymmetric model.
A simple superpotential that realises
this model is \cite{Berera:1998cq},
\begin{equation}
W =  4m_{\phi} S^2 + \lambda S^3  + \sum_{i=1}^{N_M} \left[2gM_i  X_i^2 + fX_i^3
  -2g S X_i^2 \right] \;.
\label{dmmsp}
\end{equation}
Here the bosonic part of the chiral superfield
$S = \Phi + \theta \psi + \theta^2 F$, with
$\theta \psi \equiv \theta^{\alpha} \psi_{\alpha}$ and
$\theta^2 \equiv \theta^{\alpha} \theta_{\alpha}$, is the
inflaton field $\Phi$, with $\langle \Phi \rangle = \phi$,
and it interacts with both the Bose and Fermi fields of the
chiral superfields $X_i = \chi_i + \theta \psi_{\chi_i} + \theta^2 F_{\chi_i}$.
The potential terms of the Lagrangian are obtained from Eq. (\ref{dmmsp})
by standard procedures;  the potential is
$L_V = \int d^4x d^2 \theta W(S,\{X_i\}) + h.c.$,
and the auxiliary fields $F$ and $F_{\chi}$ are eliminated through the
``field equations'',
$\partial W/\partial F = \partial W/\partial F_{\chi_i} = 0$,
which results in the Lagrangian only being in terms of the
Bose and Fermi fields.
For the above superpotential Eq. (\ref{dmmsp}), this leads
to a $\Phi^4$ inflaton potential with
interactions to the $\chi_i$ fields similar to Eq.  (\ref{dmmchiint}) and
in addition corresponding interaction terms to the 
Fermi fields $\psi_{\chi_i}$.  The distribution of the mass
scales $M_i$ are along the interval which $\phi$ traverses during the
inflationary period.
The $\Phi^4$ self-coupling must remain small
for successful inflation.  In this SUSY theory 
it occurs because the renormalization group equations
for the quartic coupling are proportional to the coupling
itself, which means even if there is another large coupling,
this will not lead to a problem.
This model can generate warm inflation with adequate
e-foldings to solve the horizon and flatness 
problems \cite{Berera:1998px}, as well as
produce observationally consistent primordial fluctuations
\cite{Berera:1999ws}.
It should be noted that the most general superpotential would
include a term in Eq. (\ref{dmmsp}) linear in the $X_i$ fields,
$S^2 X_i$. It has been eliminated by hand.
This term induces a $\phi$ dependent
mass term to all the $X_i$ fields and so it
must be very small for the success of this model.
The stability of the SUSY theory under radiative corrections
allows this term to be eliminated by hand.
A more elegant way to prohibit the linear term in the superpotential
would be by imposing a charge under some,
for example GUT, symmetry so that these $X_i$ fields are not
singlets.

More recently this model was studied in \cite{Bastero-Gil:2018yen}
for various types of mass distributions.  It was found the model
can realize warm inflation for a a wide parameter range and in
good agreement with Planck legacy data.  We also found parameter ranges
for this model entering into the strong dissipative regime,
with the inflation mass $m_{\phi}$ just over the Hubble scale.
This is not a clean solution to the swampland criteria but
it comes very close.

In \cite{Berera:1998cq,Berera:1999wt} it was shown that the DM model
can arise from
a fine structure splitting of a {\it single} highly degenerate mass level.
Let $M \approx g|M_{i+1} - M_i|$ denote the characteristic splitting scale
between adjacent levels.
{}For typical cases 
studied in \cite{Berera:1998px,Berera:1999ws}, it was shown
in \cite{Berera:1999wt} that for significant expansion e-folding, $N_e>50$,
warm inflation occurred in the interval $10^3M
\stackrel{<}{\sim} \phi \stackrel{<}{\sim} 3 \times 10^3M$ and of note, at
temperature $M \stackrel{<}{\sim} T$ and not $T$ at the much higher scale of
the mass levels $\sim 10^3M$.  The shifted mass couplings is precisely
what makes these massive states light.
In the string picture, this arrangement
corresponds to a fine structure splitting of a highly degenerate state of very
large mass, around the string scale $\sim M_{\rm string}$, with 
the fine structure splitting scale several orders
of magnitude less than the mass of the state, say $M \stackrel{<}{\sim}
M_{GUT} \sim 10^{-3}M_{\rm string}$.

The following string scenario was suggested for this 
model in \cite{Berera:1999wt}.
Initially in the high temperature region, some highly degenerate and very
massive level assumes a shifted mass coupling to $\phi$.  All the states
in this level are degenerate, so at this point they all couple identically as
$g^2 \sum_i (\phi-M)^2 \chi_i^2$.  The string then undergoes a series of
symmetry breaking that split the degeneracy and arrange the states into a DM
model $\sum_i (\phi-M_i)^2 \chi_i^2$ with $0 < (M_i-M_{i+1})/M_i \ll 1$.

In many models, thermal loop corrections are difficult to control
adequately to maintain the required
flatness of the potential and tiny inflaton mass.  
The DM model could adequately control loop corrections, 
but the interest was to find more models.
This led to developing the two stage
dissipative mechanism of warm 
inflation based on supersymmetry \cite{Berera:2003kg}, in which the inflaton
$\Phi$ is coupled to a set of heavy fields $\chi$ and $\psi_{\chi}$, which in
turn are coupled to light fields $y$ and $\psi_{y}$.  The key point is the
heavy fields are not thermally excited, which means the loop corrections to the
inflaton potential are only from vacuum fluctuations, 
and these SUSY can control.
A generic superpotential that realises the two stage mechanism is
\begin{equation}
W_I = \sum_{i=1}^{N_{\chi}} \sum_{j=1}^{N_{\rm decay}} \left[g S X_i^2 + 4m
  X_i^2 + hX_iY_j^2 \right] \;,
\label{wi2stage}
\end{equation}
where $S = \Phi + \psi \theta + \theta^2 F$, $X = \chi+ \theta \psi_{\chi}
+ \theta^2 F_{\chi}$, and
$Y = y + \theta \psi_{y} + \theta^2 F_{y}$
are chiral superfields.  The field $\Phi$ is identified as the inflaton
in this model with $\Phi = \phi +\kappa$ and $\langle \Phi \rangle =
\phi$.  In the context of the two stage mechanism, $X$ are the heavy fields
to which the inflaton is directly coupled and these fields in turn are coupled
to light $Y$ fields.  A specific inflaton potential has to
be chosen in order to assess the effect of this interaction structure
on radiative corrections. Consider
the case of a monomial inflaton potential with the additional
superpotential term $W_{\phi} = \sqrt{\lambda} S^3 / 3$ so that
\begin{equation}
W = W_{\phi} + W_I \;.
\end{equation}
At tree-level the inflaton potential from this is
\begin{equation}
V_0(\phi) = \frac{\lambda}{4} \Phi^4 \;.
\label{vophi}
\end{equation}
When $\langle \Phi \rangle = \phi \ne 0$, observe that
the vacuum energy is nonzero, which means SUSY is broken.
This manifests in the splitting of masses between the $\chi$ and $\psi_{\chi}$
SUSY partners as,
\begin{eqnarray}
m_{\psi_{\chi}}^2 & = & \left[ 2  g^2 \phi^2 + 16\sqrt{2}mg \phi + 64m^2
  \right]\;,  \nonumber  \\  
m_{\chi_1}^2  &   =  &  \left[  \frac{1}{8}  (g^2  +
  \frac{1}{2}\sqrt{\lambda} g) \phi^2 + \sqrt{2}mg \phi + 4m^2 \right] =
m^2_{\psi_{\chi}} + \sqrt{\lambda}g \phi^2\;,  \nonumber \\ 
m_{\chi_2}^2 & = &
\left[ \frac{1}{8}  (g^2 - \frac{1}{2}\sqrt{\lambda}  g)\phi^2 + \sqrt{2}mg
  \phi + 4m^2 \right] = m^2_{\psi_{\chi}} - \sqrt{\lambda}g \phi^2 \;.
\label{cpmass}
\end{eqnarray}
This implies the one loop zero temperature effective potential correction
\begin{equation}
V_1(\phi)    \approx   \frac{9}{128    \pi^2}    \lambda   g^2    \phi^4
\left(\ln\frac{m^2_{\psi_{\chi}}}{m^2}   -2  \right) \ll   V_0(\phi)  =
\frac{\lambda}{4} \phi^4\;.
\end{equation}
This is more suppressed than the tree level potential Eq. (\ref{vophi})
and so will not alter the flatness of the inflaton potential.
In \cite{Bartrum:2013fia} the two-stage
mechanism was applied to the $\Phi^4$ potential
and we showed that dissipation suppresses the tensor-to-scalar ratio.
The same behavior was observed a few years earlier
in \cite{Bastero-Gil:2009sdq},
well before the Planck data indicating the suppression of the tensor
mode.  At around this time there was anticipation based on
the cold inflation chaotic $\Phi^4$ model that a high tensor-to-scalar
ratio would be found placing inflation at the GUT energy scale.
Warm inflation demonstrated that this need not be the case for
the $\Phi^4$ model and that the presence of radiation during
inflation could suppress the tensor mode.

The above early models of warm inflation demonstrated that this type of
dynamics can be realized within quantum field theory. However these
models required a large number of fields and so were quite complicated.
On the one hand such large number of fields can be accommodated within
string theory based models, as also has been demonstrated for
the above models. Nevertheless a significant step forward we felt would
be to find warm inflation dynamics within a much simpler model.
One of the things we began to realize from developing the above
models is supersymmetry, though it can control the inflaton
potential, is clumsy to work with.  We started exploring
other symmetries that might also maintain the ultraflat potential
required for inflation.  One idea we had been thinking about was the
inflaton as a Nambu-Goldstone boson of a broken gauge symmetry.
This eventually led in 2016 to 
the warm little inflaton model \cite{Bastero-Gil:2016qru}.
In this model the inflaton field corresponds to the relative phase between 
two complex Higgs scalars that collectively break a local U(1) symmetry. 
Fermions couple to these complex scalars through Yukawa interactions 
and both set of fields satisfy a discrete interchange symmetry, essentially 
leading to an effective theory below the symmetry 
breaking scale $M \ll m_p$ 
involving the inflaton field and two Dirac fermions with a 
Lagrangian density
\begin{equation} \label{WLI_lagrangian}
-\mathcal{L}=gM\cos(\Phi/M)\bar\psi_1\psi_1+ gM\sin(\Phi/M)\bar\psi_2\psi_2 \;,
\end{equation}
where $g$ is a dimensionless coupling and
$\langle \Phi \rangle = \phi$.
The original Lagrangian is actually  written in terms of two
complex scalar fields $\Phi_1$ and $\Phi_2$ and
then these fields are represented
in terms of modulus and phase.  Thus there is no nonrenormalizable
operators in this Lagrangian.  It is just a matter of
field representation, which is convenient when the two
complex scalars develop nonzero vacuum expectations values,
$\langle \Phi_1 \rangle = \langle \Phi_2 \rangle = M/\sqrt{2}$.
For this Lagrangian 
the fermion masses are bounded from above, such that large inflaton field 
values do not lead to heavy fermions, and in addition there is a 
cancellation of the leading thermal contributions of the fermion fields 
to the inflaton's mass.  

We showed that this model can realize warm inflation,
just requires in addition to the inflaton field, two fermonic fields
and another scalar field, and for the $\Phi^2$ inflaton potential
leads to predictions for $n_s$ and $r$ consistent with 
Planck observational data.  Moreover, increasing the dissipation would
decrease the tensor-to-scalar ratio.  This model established that
warm inflation can be realized in a simple model and showed the
scenario has significant relevance to observational data.
We also obtained the strong dissipative regime for
this model in \cite{Bastero-Gil:2019gao}, thus allowing
an inflaton mass $m_{\phi} > H$, so cleanly overcoming
any infra-red or swampland problems that a light inflaton mass can
lead to.

More recently a model was 
suggested by Berghaus {\it et al.} \cite{Berghaus:2019whh},
where the inflaton
$\Phi$ has an axion-like
coupling to a pure Yang-Mills gauge group,
\begin{equation}
{\cal L}_{\rm int} = \frac{\alpha}{16 \pi} \frac{\Phi}{f}
{\tilde G}^{\mu \nu}_a G^a_{\mu \nu} \;.
\end{equation}
Here $G^a_{\mu \nu}$ is the field strength of an arbitrary Yang-Mills
group with $\alpha = g^2_{YM}/(4\pi)$, where $g_{YM}$ is the gauge coupling.
This model was named minimal warm inflation.  They showed that
for a modest coupling this led to a thermal friction and
a thermal bath during inflation.  They also showed this model could
achieve the strong dissipative regime.  It would be of interest
to develop more first principles QFT warm inflation models,
including exploring nonequilibrium dynamics
beyond the warm near thermal equilibrium models.

\section{Advantages}
\label{advan}

Before the inflation idea, 
Harrison \cite{Harrison:1969fb}
and Zeldovich \cite{Zeldovich:1972zz} had already
recognized that primordial fluctuations could be
seeds for large scale structure and noted they
needed to be scale invariant and even came up
with an approximate value for the amplitude to
be around $10^{-4}$.  Inflation then built on these
ideas to provide a mechanism for producing the
primordial fluctuations.  The goal was to
realize this from a consistent quantum field theory
model and not just symbolic scalar field potentials,
of which one can concoct many, as seen in the literature.
This main goal of inflation so far has not been realized.
Until it is, inflation remains only an interesting
idea that still needs theoretical validation.
One of the successes of
inflation is it
can realize a Harrison-Zeldovich (HZ) spectrum and improve on it
by providing a dynamical means to slightly alter the scale
invariant spectrum through introducing a tilt.  These features all
emerge from a almost flat scalar field potential that is driving
inflation. 

Inflation would have occurred at a high energy scale above any scale
for which quantum field theory has been empirically tested.
From the success of nucleosynthesis we know our understanding
of cosmological evolution is correct from the ${\rm MeV}$ scale to today.
We also know from collider experiments how high energy physics behaves
up to the $10 \ {\rm TeV}$ scale in the context of the Standard Model.
And we know there is no mechanism in the Standard Model
for realising inflation. Thus it is safe to say that if inflation
occurred, it must have been at a scale beyond where physics has
been tested.  Under these circumstances if we are trying to
build an inflation model, the first question one must ask
is what ground rules should be followed to produce a plausible
model.  One argument is build models that predict interesting,
often called smoking gun, predictions. Then if data shows such
effects, one could claim an indirect evidence for the model.
A problem with that is cosmological data shows precious little
evidence of such exciting effects.  In fact one could say
if inflation is correct, what empirical information
we are able to gather about it is quite boring.
Even by the time of the COBE data, it started becoming clear that
inflation as seen from data was likely to be boring.  
Thus a second argument is that alongside the search for better empirical data 
as already mentioned, concurrently we have 
to ask the question can we build a
realization of inflation that is consistent with everything
we have learned theoretically and confirmed empirically about quantum fields.
In the program of warm inflation this has been one of our main goals.

In order to pursue that, certain requirements have
been imposed on the inflaton field for what is considered
the ideal inflation model. One of these, which in my paper in
2000 \cite{Berera:1999ws} I called the infra-red condition, is I required that
the mass of the inflaton field should be larger than the
Hubble scale $m_{\phi} > H$.  This means that the Compton wavelength
of the inflaton field would be sub-Hubble.
All the quantum matter fields that we have
measured from colliders have masses above the Hubble scale. A
quantum matter field in which the mass is sub-Hubble scale
means it does not realize particles, certainly for
field modes less than the Hubble scale.  Our empirical
understanding of quantum field theory so far has been in terms of
a field and particle
duality.  We have no empirical knowledge what a quantum matter
field is with masses that imply super-Hubble scale Compton wavelengths,
or whether that even makes
sense.  The second requirement I imposed was that
the scale of the inflaton amplitude $\langle \Phi  \rangle$
(which we will just denote as $\phi$) should be below
the Planck scale, $\phi < m_p$.  This condition arises since
we have no knowledge of dynamics at the quantum
gravity scale (Actually we only know for sure how
QFT, as we understand it, behaves up to the LHC scale, 
so $\sim 10 \ {\rm TeV}$
but we think the next scale where our fundamental
understanding breaks down is
all the way up at
the quantum gravity scale. It could be that our understanding of
QFT breaks down at an even much lower scale than that.).
Thus we should not build an inflation
model that breaches that scale.  Some argue that the
inflaton field amplitude is not the relevant scale,
but rather the inflaton mass, which for example in the $\Phi^4$
model would be $\sim \lambda \Phi^2$. Since $\lambda$ is tiny in
inflation models, even if $\phi > m_p$, the mass itself
is sub-Planckian. However the inflaton field could also directly
couple to gravity or other fields, thus if it is of Planckian scale,
it gets into uncertain dynamics. 
In particular when $\phi > m_p$, from an effective field theory 
perspective higher dimensional non-renormalizable operators,
such as dimension six $V \Phi^2/m_p^2$, become important
and thus can ruin the flatness of the 
inflaton potential \cite{Arkani-Hamed:2003mz}.

I imposed these conditions more than two decades ago based entirely
on empirical reasoning.  Back then of course the $\eta$-problem
was known as were the effects of higher dimensional
operators.  There was belief that these issues could
be overcome with adequate model building. However
today these constraints have been realized from string
theory in the context of the swampland conditions,
which indicates a more fundamental problem in violating
them \cite{Obied:2018sgi,Ooguri:2018wrx}.
The swampland conditions are an elaborate argument built for
a very sophisticated model.  Nevertheless its final conclusion
is supported by arguments based on simple reasoning.
This is to say if a model deviates
from the regime in which QFT has been empirically verified,
thus minimally if the inflaton mass is smaller than the Hubble scale,
\[
m_{\phi} < H \;,
\]
or the inflaton field amplitude is above the Planck scale,
\[
\phi > m_p \;,
\]
then you are now entering the twilight zone of quantum of field
theory.  
The path of least resistance in
building a theoretically consistent model of
inflation is to use only the quantum field theory
as we understand it,
which minimally means to not breach these boundaries.

Whether infra-red, $\eta$, higher dimensional operators, 
or swampland problems, etc..., the writing on
the wall has been clear for decades, that having an inflaton mass
less than the Hubble scale or an inflaton field amplitude above
the Planck scale is fraught with problems.  Rather than
fight against what quantum field theory clearly has difficulty with,
it is prudent to explore an alternative approach that sits very
comfortably in quantum field theory by looking for inflation
models where the inflaton mass is larger than the Hubble scale
and where the inflaton field amplitude remains sub-Planckian.
If the inflaton mass is bigger than the Hubble scale, 
Hubble damping will have little effect in
slowing down the inflaton field.  Thus a dissipation
term (or some other backreaction effect on
the inflaton arising from particle production)
is required with dissipative coefficient
$\Upsilon > m_{\phi} > H$.  However the presence of such
a dissipative term will imply radiation production during
inflation.  This logic guided by quantum field theory
consistency leads in a natural way to warm inflation.
There are some first principles quantum field
theory warm inflation models which have been shown to achieve the
strong dissipative regime 
\cite{Bastero-Gil:2018yen,Bastero-Gil:2019gao,Berghaus:2019whh}.  
There is still
much to be done in this direction, but the evidence is
convincing that warm inflation avoids the major model building
hurdles that hamper cold inflation.

\section{Critique}
\label{critq}

Within the field of cosmology, warm inflation seems to have developed
into a rebel idea.  It was never my intention for that
to happen. As already mentioned, when I first proposed
the basic warm inflation scenario, I had little background
in cosmology and in particular inflation.  
In science, new ideas often are greeted
with interest, which has been my experience from
the different areas of physics I have worked in.
The general attitude is
no one idea can possibly be accepted
until all reasonable ideas are given full consideration.
In this respect my experience in
cosmology has been somewhat unusual, as I unintentionally
discovered.  I found that this field had a certain large group
of researchers, who advocated for cold inflation to the extent
that they seemed to have already decided
it is the right answer and had
minimal interest in considering
any alternative picture of the early universe.
Somehow they knew that their idea is exactly
what happened in that brief minuscule fraction of a second
14 billion years ago, and there is little
need for broader thinking.
The only thing this attitude has done in slowed the development of the field
and wasted resources in the process, which could better
have served research.

From my own part, I have never tried to hype up the warm inflation
idea.  
In fact most of my effort seems to have gone in looking for all
the ways warm inflation will not work.  However in the
process, a few gems of ideas have emerged that
do work, and there are a handful
of quantum field theory based models that look promising
for producing a complete first principles solution of inflation.
I hope that the scrutinizing attitude I have taken with
warm inflation has helped to keep
other researchers working on it, focused on the central
problems or to do interesting model building or
comparison of models with data.  
Often times the success of cold inflation is 
explained in terms of the number of citations and papers
it has generated. However much of the work on cold
inflation, though contains its compelling features,
also perpetuates the same ignorances or builds more complicated
theory or QFT machinery on top of the same core problems.
Science is not a democracy nor
a popularity contest. Ultimately the cold (or warm) hard truths
catch up to you. And for cold inflation, despite its over
four decades of existence and despite its simple
picture in appearance, there remain some very difficult unanswered questions
about the viability of this idea from first principles
quantum field theory. Without there being clear and unambiguous
answers to those questions, there is not much there.
The same holds for warm inflation, but it seems to be a bit ahead
in addressing the fundamental problems. 
Nevertheless warm inflation still needs to be better understood
in terms of the underlying first principles QFT dynamics
and separately in terms of the GR evolution of density
perturbations in the presence of dissipation and radiation.
Inflationary cosmology has
been around now long enough, that the age of
innocence for this field has long past. Enthusiasm
for the basic picture can no longer be sufficient
to justify its prevalence.

In this respect
the fundamental problems confronting
inflation provide ample justification to those who
have altogether given up the inflation habit and are
looking for very different solutions for early universe cosmology.
After the first minute or so of amazement 
one has at all the cosmological problems
inflation can solve in one fell swoop, here the subject is a career
lifetime later and yet there is no fully consistent, viable dynamical model
of inflation.  
In theoretical physics, inflation 
may be one of the great ideas of our time, which upon second consideration
is maybe one dare say, not so great. 
Inflation certainly generates a large number of papers and citations.
This apparent indicator of success of inflation is also its problem, in that
the idea is so vague that for almost any feature one
can imagine in the CMB data or in model building, 
one can invent some inflation model to explain it.
However this does not feel quite like the success we are typically
used to calling success in theoretical physics.
In theoretical physics, success has a more rigorous foundation,
where the theory makes
definitive prediction based on
a derivation that is widely accepted.  And herein lies
the key point, that if inflation model building was restricted to
models that were theoretically consistent, there would
be a vast reduction in possible models and possible predictions
from inflation.  Under such restrictions whether inflation really
can produce a viable model remains to be seen.
It may well be that theoretical cosmology comes full
circle and the ideas considered early on by Wheeler and Harrison
that the superhorizon primordial density perturbations were
somehow fixed by yet unknown quantum gravity dynamics, may
be the right answer.  
The great mystery of causality may reveal itself in solving the great
mystery of gravity, and that entirely may change our perspective,
especially about cosmology.
If we are unsuccessful in finding
a fully consistent dynamical model of inflation, then
we have to keep open-minded that perhaps this
physics was already fixed at the quantum gravity scale,
and the whole inflation program is wrong.  

I am being critical here just as much of warm inflation as cold
inflation.  The inflation ideas as a whole may yet prove to
be the ether of our time.  
The ether idea was of a mysterious substance that drastically alters
spacetime.  That very much also describes inflation.  
The ether idea was motivated by the best concepts of its time,
electromagnetism and stress tensor,
just as inflation today is motivated by General Relativity.
The idea was simple but implementing it led to many complications,
just like inflation.  
It is ironic that General Relativity marked
the final end of the ether idea and yet today is used to
provide the strongest argument for inflation.
Somethings never change.

The mysterious
substance in the case of inflation is vacuum energy.  Within the GR
equations, such a substance leads to an exponential expansion,
which is characteristic of inflation.  However there has been no
direct detection of such a substance exhibiting anti-gravity.
Within conventional
QFT as we so far understand it, the vacuum energy up to a constant
is arbitrary and
fundamentally unnecessary.  Thus from this perspective,
conventional QFT can have an arbitrary amount or not of inflation,
it's completely unconstrained.  

For these reasons one needs to be very careful in working with inflation.
It should not be viewed as a goal that whatever is the evolving
data and theory, at all costs a model of inflation must exist.
Rather it should be question, a matter of scientific enquiry, that can
we find a sensible and theoretically consistent inflation model in line with
the data.  And if we can not, then there is no validation of it. If
that becomes the case, then
serious thought needs to be given
to alternative ideas about the early universe beyond inflation.

At least if one has a QFT model
of inflation which is otherwise consistent, thus
minimally  with $m_{\phi} > H$
and $\phi < m_p$, then such a model has only one unknown fundamental quantity,
this vacuum energy. And if a tensor mode is found in the CMB, 
indicative of a vacuum energy, it could then more uniquely
be attributed to a QFT model.  If on the other
hand the QFT model requires other unknown fundamental assumptions
such as sub-Hubble masses or super-Planckian field
excursions, then there are too many unknowns for the model
to be uniquely predictive and it leaves open a bigger
range of interpretations about what has been found in the CMB data.

For those of us
who work on warm inflation, the alternatives are acknowledged.
We have approached warm inflation as just one relevant idea that
needs to be fully vetted.
Suppressing alternative ideas or not fully
recognizing the success of alternatives is not
helpful to the development of this field.  Point in fact,
everything found in the CMB data to date that is talked
about in support of cold inflation,
equally is a success for warm inflation,
yet this point is rarely acknowledged.  Along with
this, today it is often
stated as a general fact that the $\Phi^2$ and fairly closely the $\Phi^4$
models of inflation have been ruled out due in particular
to the lowering of the upper bound on any possible
tensor-to-scalar ratio, but in fact such models
are still consistent within warm inflation
\cite{Bartrum:2013fia,Bastero-Gil:2016qru,Bastero-Gil:2019gao,Bastero-Gil:2018uep}.
These have
been the models for three decades that the advocates of cold
inflation had been pinning their
hopes on. Yet when they were ruled out, rather than
even a brief moment of reflection and reassessment,
thus noting the continued success of these models within warm inflation,
their interest immediately turned to other more
exotic models. What's going on?

I am not saying cold inflation is wrong.  How do I know. Everything
I know is based on what I learn from the data and theory, and so
far both these are inconclusive.  There is no one who knows more than that.
We don't have a Gandalf here showing us the way.
Our only guides are the theory and experiment.
Although there are pockets of success for inflation,
they should not be exaggerated, since the whole
picture doesn't quite add up neither from the side
of theory nor observation.
Everyone is capable of thinking for themselves in seeing the
the science is inconclusive about inflation, 
and even more about warm versus cold.
However warm inflation does
score a bit higher than cold both in having anticipated a
lower tensor-to-scalar ratio, which could still be found,
and in addressing the main model building problems.
There is no principled argument for focusing exclusively
on cold inflation.

The introduction of the Planck 2018 data was a watershed
moment \cite{Planck:2018jri}, where one would have to conclude that
there is no longer a benckmark paradigm of the early universe.
Cold inflation can no longer claim that mantle.
Even if any researcher still wishes to remain ignorant about
warm inflation or the many other interesting alternative
ideas about the early universe, the quantum gravity possibility
still hangs over the whole subject.  Thus inflation has a very
tough requirement to come up with a fully consistent
dynamical model.  That is why for warm inflation, as I
already mentioned, I have maintained very stringent requirements
in building a model that is fully consistent with
the quantum field theory we know to have worked at
tested collider energy scales.   
As such this minimally means
no scale in the model, such as the inflaton
field amplitude, should be larger than $m_p$, since that enters
the quantum gravity regime, for example in generating
nonrenormalizable operator corrections.  In going above the $m_p$ scale,
one must already make an assumptions that quantum gravity effects
somehow are subdominant to the low energy model.
However as soon as any assumptions are required about quantum gravity,
it isn't much of a leap to
simply assume quantum gravity solves all the problems.
Though I have
not achieved the complete goal up to now
of coming up with such a fully consistent
warm inflation model, I have at least shown that warm inflation can
avoid the most ambiguous regimes of quantum field theory
by models remaining where
\begin{eqnarray}
m_{\phi} & > & H , \ {\rm and}
\\
\phi & < & m_p \;.
\end{eqnarray}
Thus I have shown warm inflation
has the basic ingredients to solve the most
fundamental and pressing problems of inflation model building.
Moreover, with collaborators we have developed the first simple
quantum field theory model that can realize these properties
and give an observationally consistent result 
for inflaion \cite{Bastero-Gil:2016qru,Bastero-Gil:2019gao}.
These properties arise as consequences ultimately of the 
particle production from 
interactions amongst quantum fields during inflation.
As an aside, note these effects differ from the coupling of matter fields
to classical gravity that then leads to particle
production arising from an expanding background
and/or curvature, which
was examined in the seminal work of
Parker \cite{Parker:1968mv} and subsequently by others
\cite{Zeldovich:1971mw,Hu:1973kq}. Some papers also discussed
this in the context of inflation
\cite{Ford:1986sy,Frieman:1985fr,Cespedes:1989kh,Lyth:1996yj},
although these effects showed no appreciable change from cold
inflation to the large scale predictions.

There is a more general point here that my
work has demonstrated. Just as radiative and thermal corrections
affect the background scalar field physics by altering its
effective potential, the warm inflation lesson is
that when time evolution is involved,
additional QFT effects will occur, such as
particle production, and these effects will
react back on the background scalar field and alter its
dynamics, such as by possibly allowing a larger inflaton mass or affecting
the extent of the background inflaton's
field value.  
This broader lesson from our work
may be useful for other early universe models, to
not ignore particle production from QFT interactions
and the associated effects that come with it.
For inflation, accounting for this
from the time evolving
background scalar field could be the missing ingredient
that has balanced inflation models to
allow for more sensible model building.

My above critical comments about the attitudes in the early
universe cosmology field
arise from direct experience in trying to
develop warm inflation.
Although there
have been many researchers who have contributed
their insights and skill into developing warm inflation, it has also
faced an almost stauch lack of acceptance from that larger portion of
the cosmology community that advocated for cold inflation.
Mainly they have just minimized acknowledging the idea at all
but also argued that the idea is in someway or another inferior.
At the start their arguments centered around the Yokoyama and Linde
work, despite us having shown fairly soon after that
warm inflation could be realize from QFT.  Once such arguments
clearly became dated, others followed, 
such as warm inflation is not as simple as cold inflation since it
requires an additional parameter, the dissipation coefficient.
Cold inflation also needs an additional parameter
for reheating.  However in that scenario, it seemed to have been
decided without any empirical evidence supporting the picture,
that inflation is synonymous with a supercooled phase 
and that reheating must be a distinct separate phase.
By now, though criticism of warm inflation is much more muted,
with some past critics even working on the idea,
the main defence I have heard 
by some of the cold inflation advocates is they 
believe in cold inflation or like it.
Belief is a fine quality in a devotee but needs to be tempered
by the facts in a scientist.

The simplicity argument, despite it obvious shortcomings, carried
on being used until the Planck data showed there was no
tensor mode at the high energy scale predicted by the single field
$\Phi^2$ and $\Phi^4$ chaotic cold inflation models.  
It became further obsolete with the swampland work
that showed that many cold inflation models may not be consistent
with quantum gravity, at least within the context of string theory.
On the other hand, both these developments worked in favor
of the warm inflation scenario.  We had seen from model
calculations as early as 2009 \cite{Bastero-Gil:2009sdq}
and further studied in \cite{Bartrum:2013fia}, 
that warm inflation suppresses the
tensor-to-scalar ratio, including in the simple monomial
models such as $\Phi^2$ and $\Phi^4$.  Moreover the swampland
conditions showed once again that an inflaton field with
mass $m_{\phi} > H$ and $\phi < m_p$ 
would be the consistent regime in this case in the context
of string theory.

The Planck 2018 inflation paper \cite{Planck:2018jri}
really demonstrated the harmful degree of
advocacy by the cold inflation enthusiasts.  I had informed the
Planck inflation team about the successes of warm inflation in demonstrating
well before their data that the tensor-to-scalar ratio
should be suppressed, even in the simplest monomial models.
Given that the outcome of no tensor mode detected down to the GUT scale
was the big result from their analysis, thus breaking decades
of expectation from cold inflation of a confirmation of
the simple monomial models and possibly even its consistency condition,
it is a rather big deal that warm inflation all along had
what was now the consistent picture with the data.
Despite this information available to them, the Planck 2018 inflation
paper reported amongst its {\it ``main results''} summary
that Planck 2018 {\it ``strongly
disfavors monomial models''}, which in truth only applies to
cold inflation models, but the wording with no other qualifiers
suggests that
no science exists beyond that picture.
That is misleading.
In fact one significant interpretation of the Planck 2018 results is that
it has found indirect evidence for radiation and dissipation
during inflation, that is suppressing the tensor-to-scalar ratio.
This is a very simple explanation to the Planck 2018 data. It is the sort
of bread-and-butter explanation, involving simple potentials,
particle production, dissipation, etc... that has a familiarity.
It offers an explanation of this remote phenomenon that one ideally
wants in terms of relatable analogies. This is
a far cry from cold inflation, where the explanation involves
string theory, modifications to gravity and other features that
so far have shown no close contact with reality.
Physics is an empirical science. An unknown idea used to
understand another unknown idea gains you very little.
Neither string theory nor modifications to gravity have been shown
to be theoretically consistent, and there is no experimental
evidence for them. Thus it's unclear how relevant any
inflation models are that are based on these ideas.
On the other hand, models based on the quantum field theory
we are familiar with, no doubt extending beyond the Standard Model
but following the basic rules that are understood to work,
have closer connection with reality. In this respect there are
at least a few warm inflation models.

Nevertheless Planck 2018 gave
only the vaguest mention about our successful
prediction, yet
the associated papers going back to 2009 that had made this
prediction were not referenced.
As Planck are an observation group,
their responsibility is to their data
and protecting its integrity from any theoretical bias.
A supporting pillar of science is impartiality of experimentalists
to theory.
So if they insist on discussing theory,
then they need to be equitable to all ideas that predicted the trends
in their data.
In this respect warm inflation has been spot on
and with a very simple explanation,
yet was given only a vague mentioned in their paper with
relevant references missing.
It does not benefit any
field of science when objectivity is lost and
advocacy reaches the point of discarding
the evidence on the ground, with facts not thoroughly reported.
And here the facts remain that warm inflation bucked
the theoretical trends by arguing for a lower tensor-to-scalar ratio below
the detection limits of Planck, and we were correct.

If a tensor mode is eventually detected at some
lower energy scale, then there are very good claims
to be made that this is evidence for warm inflation
\cite{Bartrum:2013fia,Bastero-Gil:2016qru,Bastero-Gil:2019gao,Bastero-Gil:2018uep}.
Statistically warm inflation models compare very well to
Planck data. The $\Delta \chi^2$ for warm inflation
with dissipation of the type found in the warm little and
two-stage models are much smaller then typically found
for cold inflation models.  For example for the $\Phi^4$
potential, the $\Delta \chi^2$ for 
warm inflation models is very small, order $1$
\cite{Benetti:2016jhf},
whereas for cold inflation was
found to be $\sim 40$ in the
Planck 2015 analysis \cite{Planck:2015sxf}.
Moreover, for the warm little inflaton model
in \cite{Bastero-Gil:2019gao}, for the
$\Phi^2$ potential we found that a super-Hubble scalar inflaton
mass and sub-Planckian scalar field excursion throughout inflation
occurs for a tensor-to-scalar ratio $r \approx 6.4 \times 10^{-6}$
and with a spectral index $n_s \approx 0.965$ so within
the $68 \%$ confidence level of the Planck 2018 
legacy data \cite{Planck:2018jri}.

Imagine for a moment that tomorrow a tensor mode is found in the CMB,
which corresponds to an energy scale for inflation just below
the present upper bound (which itself is just below that 
found by the Planck 2018 results).
Then if we go by the Planck 2018 inflation paper \cite{Planck:2018jri}
{\it ``main results''} summary,
a likely best fit to that hypothetical result is D-brane inflation.
So does that mean that will mark the day
when string theory will have been experimentally
discovered?
If we go by their main results, they say
{\it ``...inflationary models such as $R^2$, T and E 
$\alpha$-attractor models, D-brane inflation and those having
a potential with exponential tails provide good fits...''}
to their data.  The first three of these 
involve either modifications to gravity, supergravity, conformal or
superconformal field theories, and/or string theory,
so in all cases would be very major discoveries.
Such models are important as intellectual exercises in the overall
efforts to develop more fundamental theories of physics.
However they are far from established theories 
of the physical world. Such models have no urgency for an accurate best fit
comparisons by experimentalists against their data
and for them to do that is misguided.
The last possibility they gave of potentials with an exponential
tail is rather disappointing to find in their
list. This suggests that after four decades of inflation model building,
they consider it a main results to give just a form of a potential,
which moreover without a higher theory that might justify it,
on its own is nonrenormalizable and so without merit.
It only should be added that this potential is much more 
complicated than the simple, renormalizable,
monomial potentials that work for warm inflation.
All their main result possibilities are more complicated to
the  much simpler conclusion, that their
CMB data implies an inflation with a simple, conventionally renormalizable,
potential and just a bit of radiation and dissipation.
It would be irresponsible and downright unscientific to
favor the complicated explanation over the simpler one.
This goes against all practice in science in which when given the option
to take the simplest interpretation.
That is the misdirection the Planck 2018 inflation paper is
heading us toward, if we simply ignore the warm inflation possibility.
Out of over fifty inflation models tested (this in any case seems excessive),
they did not test any warm inflation models,
which are amongst the only few that are conventionally
renormalizable, thus by default significant.
There are only two dynamical pictures of inflation, warm and cold.
If a thorough scientific analysis is the goal, then models
from both pictures need to be tested against data.
The particle physics dynamics implied by an interpretation
of the CMB data as either
warm or cold inflation is very different. Thus there needs
to be extreme caution
in how data is interpreted,
as it will become the underlying basis for future
particle physics model building.

Without a rigorous model of inflation that is fully
consistent with the QFT that we know and understand, it's even
dangerous to interpret any hypothetical tensor mode discovery
in the CMB as inflation in the way we currently understand it.
There is still a possibility that the true explanation may involve
quantum gravity in some as yet subtle way. For example
such a tensor mode may arise from an inert vacuum energy at
the measured scale, but the primordial density perturbations
may still have been fixed by quantum gravity. Moreover there may
be some way the inert vacuum energy is correlated with quantum
gravity.  It may also be that quantum gravity is creating
tensor modes of a sort we don't fully comprehend at the moment
and we may confuse those signals for a vacuum energy.
All this may sound like contrived explanations, but we
just don't know all the unknown unknowns about quantum gravity
until it is fully solved.  That is why
it is imperative that if we want to interpret a hypothetical
tensor mode discovery in terms of inflation as we understand it,
then minimally we need a rigorous QFT model 
that tells us how exactly do we understand inflation. 
And this model needs to be not half rigorous
or sort of rigorous but completely rigorous.  This
requirement is true for both warm and cold inflation.

In fact, in the fallout of Planck 2018 it is
best to not bias our thinking just toward inflation,
especially by CMB experimental groups.
If any experimental group insists on discussing theory,
then they need to be very disciplined in giving a fair
and broad overview.
Experimentalists should understand what they
represent as experimentalists if they
make any association between their results and string
theory or any of these many speculative ideas from theoretical
physics.  Actions such as these shift the focus away
from looking for results that
might be accessible to results which, for the time
being, are impossibly unreachable.
One expects experimentalists,
especially large experimental groups,
to lean on the side of restraint in
associating any theoretical idea with consistency
to their data until they have confidence the data 
is headed with high likelihood eventually to confirm that idea.
They need to think very carefully whether the fragmentary
information gleened about the early universe from the culmination
of all their extensive work really is enough, to start associating
it with these fantastic ideas of theoretical physics.
And if they insist on wanting to do this, they need to ask
what kind of credibility are they giving their results if
they then ignore simpler, more tame, theoretical explanations,
by not giving a broad assessment of theory.
Very speculative directions can be left for individuals to
consider, but experimental groups should act responsibly
toward the whole field and also
make thoughtful use of hard earned taxpayer money. 
Actually, there is no need for experimentalists to get into any
comparison of theoretical models. 
Once their data is out, within a short period
some theorist will analyze it and report that some string
theory model or whatever is the best fit to data and so forth.
But that's different. Nobody listens to us theorists when we
talk like that.  However for an
experimental group to make
such claims is a huge deal.

The discussion in this Section has shown that at the very least
there are two quite different dynamics, warm and cold inflation,
with models which agree very well with CMB data.  Moreover this is
likely to still hold if a tensor mode is discovered in the near
future.  So there already is a large degeneracy of possible
early universe solutions, even before considering
non-inflationary models.  We are far off from any conclusive
early universe theory.  If a tensor mode is discovered,
more needs to be assessed whether to present understanding
its explanation reduces just to
a comparison between warm and cold inflation.  Thus,
further research should be done in searching for non-inflationary
early universe models that create tensor mode 
signals \cite{Brandenberger:2006xi},
as well as more scrutiny over secondary sources of 
such B-mode polarization signals.  
Alongside this, as discussed in the next Section,
attention should be given to the level
of assumptions entering a model, with those requiring the least
number of assumptions then being most important for
comparison against data.
At this point it is to our advantage to treat all viable
ideas about the very early universe on an equal footing
and give them all equal tests in comparison to the data.
In the long run this approach is better also for cold inflation,
so that it is fully vetted and scrutinized.
We need to avoid the danger of talking ourselves into believing
only one idea is correct.  A broader perspective is needed until
clear evidence
favors one idea well above all others, including
the possibility that none of our present ideas for now are adequately favored.

\section{Model selection}
\label{modselect}

Apparently Landau once said something to
the effect that cosmologists are often in error
but never in doubt.  
I don't know if he said this but it sounds relevant.
Still today this statement holds meaning for the theoretical side
of cosmology.
The problem with theory in cosmology is that no matter what particle
physics model one builds, only a small portion will directly be tested
against what intrinsically is limited empirical information about the early
universe.  In particular when
comparing to data, all inflation models boil down to
the value of a scalar field potential at one particular
point of the background field and a couple of derivatives
at that point (and maybe a couple more).
The lower multipole spectrum can also be fit covering
up to around 10 e-folds of inflation, so giving partial
information about
a small region of the potential about this point.
There are a few more details. The potential in question must
have another point at which inflation ends, and the
point of interest for comparing to data is meant to be
from where the inflaton evolves long enough to create
around 50 efolds of inflation. Moreover the end temperature also
gets fixed by the inflaton model and that is involved in
determining exactly how many efolds of inflation near about 50 will
be needed.
For warm inflation there is also the radiation density
generated during inflation, which in the most common
realizations is a temperature scale.  A point
on the scalar potential will be associated with a
temperature, which itself will depend on the parameters
of the potential and some set of interaction couplings in the model.
So the parametric dependence from the underlying model
is a bit different in warm versus cold inflation, but
the basic idea is the same.

Details aside, inflation
model building boils down to finding an encasing theory that
can produce a scalar field potential with a point
on that potential and a small region about that
point that compares well against the data.
If that point on that potential of that encasing
theory agrees well against the data, that does not
mean you have confirmation of the encasing theory.
It is the other way around. You need to show that the
encasing theory which produces the potential that has
that point has some claim to be a physical theory.
For example, the presence of the photon does not 
mean string theory has been discovered.
String models contain the photon, but the challenge is
to show a string model also is theoretically consistent.
If there was a unique encasing model that produced
the point on the potential which agreed with 
the cosmological data, then that would
at least single out that model. 
However, what compounds the problem is the proliferation of
inflation models in the literature, which suggests not
to expect such a unique association, and also
not all these models ultimately could be 
based on underlying correct theoretical ideas. 
Moreover for many encasing theories,
the choice of the inflaton potential
has some degree of arbitrariness, thus further dissociating
the potential, which is the only part of the encasing theory
that cosmological data is testing, from other aspects of the encasing theory. 

The cosmological data
provides an upper bound on the tensor-to-scalar
index $r$, an amplitude of the scalar
perturbation, and the spectral index $n_s$ and possibly its running.
The parameters of a given inflation model need to fit as best they
can to these observables alongside some constraints such as producing
a sufficient duration of inflation, the final temperature
after inflation, and perhaps some theoretical constraints
and consistency conditions within the model.
The error in the spectral index is sufficiently wide that
one should not anticipate uniquely separating out a single 
inflation model. Cosmic variance is an ultimate limiting
factor in how accurately $n_s$ can be measured.
Narrowing these errors will
certainly improve predictability, but more
is still needed to help determine the best model.

The requirement that the inflation
model be theoretically consistent
and have a claim on being part of the physical world provides
a useful guide.
It suggests that another helpful measure for separating inflation models
would be to characterize for each such
model how speculative it is.
The more speculative it is, the higher the chance
that once quantum field theory and particle physics model
building is better understood, the prediction from
the model will very likely change and in the worst case the underlying ideas
the inflation model is built on are wrong or inconsistent.
Alternatively, the less speculative the encasing theory is the more
predictive it is.  The least speculative theory would be
the Standard Model, but we know that is not adequate
to explain cosmology.

This gives a guiding rule, that
models of the early universe should be assessed 
on how much more beyond the SM is required
to compare well against 
the cosmological data.
By counting the number of speculative ideas
inputted to a model, it
would provide some
guidance over model building.
At the lowest end of speculation it would
still require building an extension to the Standard
Model but remaining within the rules of QFT that
have been understood by the Standard Model.
At a bit higher level in speculation, effective field theory
methods, though less predictive, can still be OK provided
the cutoff scale is below the quantum gravity scale.
Its fine to still make much more speculative models at a
much higher end of the speculation count, but
by having such a count, it makes us
more aware that such models are
for the most part intellectual exercises.  The models
with the lowest speculation count are the serious
contenders for comparing to data.
In that respect
models that
involve any assumptions about quantum gravity or
rely in an essential way on some modification to gravity
are not at a stage where much is gained to compare them to 
the very limited data that cosmological measurements can provide.
That is because first our ideas about quantum gravity 
or even some aspects of classical gravity beyond GR could
and probably are wrong at some level 
at the moment, so any model based on them
will most likely ultimately need adjustments or
simply be wrong. Second,
if one is ready to make such assumptions 
involving quantum gravity as part of
their model, then it's not a 
step much further to simply assume that quantum gravity might
completely solve the problems of the early universe 
by directions already suggested in the literature or
in some entirely different way, once
it is much better understood.  Thus for any
such model, we simply
must wait until quantum gravity is much better understood before we can
assess its relevance.
These are points I hope our research over the years in warm inflation
has tried to express through our actions in attempting 
to find a warm inflation
model that is fully consistent with the quantum field theory
as we presently understand.  Admittedly, we have also crossed into
the higher end of the speculation count, because it 
has fundamental interest and is fun to do,
but at the same time we have put considerable effort to find
much more experimentally relevant models 
that were as close an extension as possible to the QFT we
presently understand.  Here I will try to a provide a more
systematic guide as to how to do a speculation count
for any cosmological model.

\medskip

\begin{center}

\begin{tabular}{l}
\hline
\bf{Table: Range of speculative attributes in cosmological models} \\
\hline
\bf{Fundamental [F]} \\
\hline
Quantum gravity \\
Additional spacetime dimensions above four \\
Modifications to gravity beyond General Relativity \\
Sub-Hubble mass scalar fields \\
Supersymmetry/other new spacetime symmetries or adjustments to them \\
\hline
\bf{Technical [T]} \\
\hline
Effective field theory methods with cutoff scale below $m_p$\\
Symmetries included in the model not of the type in the Standard Model and excluding new spacetime symmetries \\
Symmetries included in the model \\
Extra fields added beyond the Standard Model and not attributed to any symmetry \\
Model building beyond the Standard Model \\
\hline
\end{tabular}

\end{center}

\medskip

Quantum gravity, string theory, higher spacetime dimensions
beyond four, modifications to spacetime
aside from additional dimensions, loop quantum gravity,
super-Planckian field excursion, higher dimensional operators,
sub-Hubble masses, modified gravity,
fields where any mode propagation differs from 
standard QFT,
supergravity,  
supersymmetry, effective field theory with cutoff scale
at $m_p$, using effective field theory methods,
extra fields beyond those in the Standard Model which can not
be attributed to some symmetry or higher theory,
symmetries not of types in the Standard Model,
symmetries included in the model,
model building beyond the Standard Model, etc...,
each add more speculation to a model.
To account for these properties,
the speculation count would need two separate categories.
One would be for fundamental (F) attributes that alter the quantum
field theory as we presently understand it.
These are all attributes with unknown unknowns, for
which at the moment there is no
empirical evidence for whatsoever, and alongside
that there is no established theory for them.
Inflation models with any fundamental speculation counts are susceptible
ultimately to being wrong or needing significant
modifications once any of the fundamental
attributes in the count are better
understood in the future.
The other category
would be speculations of a technical (T) nature related
to standard QFT model building.  These are
attributes with known unknowns or known knowns.
The success of the Standard Model is strong evidence that
standard QFT model building is immensely successful,
with ample evidence of its empirical relevance.
The rules for such model building were set decades ago and by now
are prescriptive, leaving much of it as a technical exercise.
It can still lead to novel results,
but it works with symmetries and properties of the type
familiar or similar to those in the Standard Model.
The attributes belonging to both categories are given in
the Table.
The speculation count based on this Table applies to inflation
models that have a mechanism to end inflation into the radiation
dominated regime and are able to protect the flatness needed
of the inflaton potential from radiative/thermal corrections.

The separation into two different categories is necessary because it
is an apples and oranges type comparison of speculation between them.
Attributes in the Fundamental category involve a conceptual
leap beyond our present both theoretical and empirical
knowledge. On the other hand, for models with all 
attributes in the Technical
category, they have a familiarity based on our
extensive experience with the Standard Model.
Models with only Technical attributes
are built on types of quantum fields, all that have
been tested in collider experiments, and so there is a possibility
that beyond cosmological data some particle physics
based collider or astrophysical tests
can also be conceived to test such models, even
if indirectly.  To appreciate the distinction between
Fundamental and Technical attributes, suppose
a new particle was discovered which could be explained
through standard type of model building beyond the Standard Model.
That would be exciting and extremely noteworthy, but mainly
because any discovery in particle physics has
basic significance, happens slowly, and
after a lot of work.
Yet it would be a level shift fundamentally higher if
for example the discovery directly showed that our world had an extra dimension.
The Fundamental/Technical speculation count is obtained by
simply counting the attributes of a model in each category.

The fundamental count includes sub-Hubble mass scalars based on
my discussion earlier in the paper. Models with super-Planckian
field excursions or higher dimensional operators 
(not from an effective field theory with cutoff below $m_p$), 
would include quantum gravity in the fundamental count.
Even if quantum gravity is not explicitly used,
once a model has these features, implicitly it is affected
by quantum gravity and that is an
uncontrolled approximation.
In the technical category, any symmetry included in the
inflation model is counted to assess the complexity of
the model, but if that symmetry
differs significantly from types found in the SM, it
is counted twice to account for the higher
speculation associated with it.
For symmetries much different from the SM, specifically I
exclude any new spacetime symmetries, like supersymmetry or
alterations to Lorentz invariance,  CPT etc...,  as that
is already counted in the Fundamental category,
but it would include for example technicolor, preon models, unparticles etc...
and for theories in the Fundamental category include
symmetry details such as choice of compactification, brane type etc...
The inflaton potential, which is often arbitrary in inflation models,
I did not include as a separate attribute.  This is partly
accounted for in the model building beyond the SM attribute.
Also from the symmetries in an inflation model, usually
a scalar emerges that is identified as the inflaton. If the potential
is nonrenormalizable with cutoff scale below $m_p$, it is also
accounted for with the effective field theories with cutoff below
$m_p$ attribute and if the cutoff scale is at $m_p$ it is accounted
for with the quantum gravity attribute.
Counting the number of speculations
in a model mitigates the need for relying on individual 
opinions on this matter.
Two categories are necessary since
Fundamental attributes are a different degree of speculative to the Technical
ones and it would be impossible to give any metric
to compare between the two. Then within each category,
a priori with no further information,
the count treats
all attributes equally
and the degree of speculation in a model
is simply down to how many of the attributes 
in the Table it has.

One thing we can all agree on as theorists is a particle physics model
of cosmology will require some degree of speculation,
but we will never agree amongst us the degree to which
each of the possibilities in the Table is speculative. 
This speculation count provides a simple step, of at least 
counting each speculative
attribute.
What we will still
not agree amongst us is what level of
speculation is acceptable, since speculation might also be viewed
by another word, insightful.  However there are
two limiting cases on which we can all agree. First,
models involving quantum gravity
are at the extreme end of speculation.  
We know so little about this thing we so conveniently call
quantum gravity, that
we don't even know if near the Planck
scale, physics behaves by the rules of quantum mechanics
or whether at some scale, possibly much below the Planck
scale, some entirely new type of physics kicks in that
supercedes quantum mechanics just as that supercedes classical
mechanics at around the atomic scale.
Gravity is the only force, which interacts directly, as far as we know,
with all forms of energy and defies being
encaged as a renormalizable point particle quantum field,
so befuddles attempts at a unitary theory.  These properties
that separate gravity from all the other fundamental fields of Nature
may be the
earliest hints that at some high enough energy scale,
its behavior goes beyond not four spacetime dimensions
but rather beyond the rules of quantum mechanics.  
At present no one can exclude
that possibility, thus there is little point in
trying to argue that models requiring quantum gravity
have any unique or
urgent phenomenological relevance.  In fact,
we don't even have any definite idea how physics behaves just a
couple of orders of magnitude in energy above the LHC scale,
and so the Planck scale is well-beyond reach for
meaningful application to phenomenology.
Second,
for any model at a low level of speculation, there is
much less to be theoretically debated, so 
there is greater significance in testing how well it compares to data.
In truth we don't even have a definite
idea what is just around the corner at the next higher energy 
above the LHC scale.  So even cosmological models with low speculation
count should be treated with a great deal of caution.
Nevertheless, for any model with a low level of speculation and
especially with no Fundamental attributes, if it also  
fits well against data, then on a relative comparison it is
amongst the best cosmological models.

The higher the speculation count, especially
in the fundamental category for a model, the more
it presses the question whether
in order to obtain an adequate
phase of quasi-exponential expansion, is the proposed model really
a measured solution.  For example is adding six spacetime dimensions 
really a measured solution to just obtain a phase of inflation.
The count forces a think about the purpose
of a model. If one is developing string theory, it makes sense
to see how well a string based model can realize phenomenology relevant
to real world data.  However if one is interested in determining
the most relevant models that agree with the data, too high a
speculation count indicates those are not the primary models
that should be tested.  Cosmological observation basically
fixes two data point, the scalar amplitude and the scalar index $n_s$ 
and gives one
bound, on the tensor-to-scalar index $r$, and maybe
a few more data points
such as nongaussianity, running of $n_s$, isocurvature etc...
As the speculation count rises, the assumptions going into a model
swamp the limited data and nothing is really being tested

\bigskip

Here we examine the Fundamental/Technical speculation count for a few models.
We are not concerned here about how well the models compare to
data, which we have already discussed in previous
parts of the paper.  Here the count is just assessing
the theoretical aspects of these models. The potentials
are written in terms of only the background inflaton
field $\phi = \langle \Phi \rangle$.

\medskip

\noindent
{\bf D-Brane inflation model \cite{Dvali:2001fw}:}
D-branes are solitonic solutions arising in 
string theories of type I, IIA and IIB.
There is an interaction energy between two parallel
brane and anti-branes, and this is the potential
energy utilized to drive inflation.
The inflaton field is a mode corresponding to a relative motion
between two parallel branes.
The model
relies on the locality of the higher dimensional theory
to allow for a sub-Hubble mass as necessary in cold inflation.
For D-3 branes, the potential has the form,
\[
V_{\rm D-brane}(\phi) = M^4 (1-\alpha/\phi^4) \;.
\]
The counting of speculations from the Table
entering the D-brane inflation model 
is below.  Here the square bracket indicates whether
the attribute is Fundamental [F] or Technical [T], and
the curved bracket gives the number of speculation counts
for that attribute if it's larger than one:  
\begin{itemize}
\item[-] Quantum gravity [F]
\item[-] Dimensions beyond four (6) [F]
\item[-] Sub-Hubble mass inflaton [F]
\item[-] Supersymmetry [F]
\item[-] Symmetries not of type in SM - choice of compactification and D-p brane (2) [T]
\item[-] Symmetries included in the model (2) [T]
\item[-] Model building beyond the SM [T]
\end{itemize}
This gives a speculation count for Fundamental/Technical
properties of 9/5.  Here the choice of compactification and D-p brane
I include in the technical category and not also as a fundamental
attribute for new spacetime symmetries, since this model
already has been penalized in the fundamental category for
extra dimensions, which is sufficient.

\medskip

\noindent
{\bf $\alpha$-attractor superconformal inflation
model \cite{Kallosh:2013yoa}}:
This is a supergravity model where 
the parameter $\alpha$  is inversely proportional to the
curvature of the inflaton K{\"a}hler manifold.  
A common choice of
potential is
\[
V_{\rm \alpha-attractor}(\phi) = \tanh^{2n}\left(\frac{\phi}{\sqrt{6\alpha}}\right)\;,
\]
for $n,\alpha > 0$.
For large 
curvature, which corresponds to small $\alpha$, the predictions
agree well with CMB data.  
The counting of speculations entering this $\alpha$-attractor
model is:
\begin{itemize}
\item[-] Quantum gravity [F]
\item[-] Sub-Hubble mass inflaton field [F]
\item[-] Supersymmetry [F]
\item[-] Symmetries not of type in SM - superconformal [T]
\item[-] Symmetries included in the model - three chiral multiplets and 
K{\"a}hler potential with superconformal and $SU(1,1)$ symmetries (5) [T]
\item[-] Model building beyond the Standard Model [T]
\end{itemize}
This gives a speculation count for Fundamental/Technical
properties of 3/7.


\medskip

\noindent
{\bf $R^2$ Starobinsky model
\cite{Starobinsky:1980te,Vilenkin:1985md}:}
This is a type of modified gravity model which
has a curvature-squared $R^2/(6M^2)$ term added
to the Einstein-Hilbert action, where $R$ is the
Ricci scalar and $M < m_p$.  
This action is transformed into the Einstein
frame leading to a inflaton potential of the form,
\[
V_{R^2}(\phi) = \Lambda^4 \left[1 - \exp\left(-\sqrt{\frac{2}{3}} \frac{\phi}{m_p} \right) \right]^2 \;.
\]
The counting of speculations entering the $R^2$-Starobinksy 
model is:
\begin{itemize}
\item[-] Quantum gravity [F]
\item[-] Modifications to gravity beyond GR [F]
\item[-] Sub-Hubble mass inflaton field [F]
\item[-] Symmetry not of type in SM - transform from the Jordon to Einstein frame [T]
\item[-] Symmetry included in the model [T]
\item[-] Model building beyond the Standard Model [T]
\end{itemize}
This gives a speculation count for Fundamental/Technical
properties of 3/3.
In the original paper by Starobinsky, he had viewed the $R^2$ term
as dynamically generated
as a self-consistent
solution of the vacuum Einstein equations by one loop
corrections due to quantized matter fields.
The model can also have quantum gravity interactions treated semiclassically
but these become subdominant for sufficient number of matter
fields.  Thus one could also count the assumptions from such a more
first principles approach, but
that would need the details
about the matter fields and interactions. 
Nevertheless in such a case
the fundamental assumption added in our above list of
modifications to gravity beyond GR
would not be included, although assumptions about the underlying
matter fields would need to be added.

\medskip

\noindent
{\bf  Higgs Inflation model \cite{Bezrukov:2007ep}:} 
This model assumes there are no
other fields in the universe aside from those in the Standard Model,
and the Higgs field has a non-minimal coupling to gravity.
In the initial Jordan frame the Higgs field, $h$, 
has a standard type of
quartic symmetry breaking potential of the form
$\sim \lambda (h^2 - v^2)^2$. To get rid of the non-minimal coupling
to gravity, a conformal transformation is done to the Einstein frame.
The Higgs field is then treated as the inflaton, which at high field value 
has the potential
in the Einstein frame,
\[
V_{\rm Higgs \ inflation}(\phi) = \frac{\lambda m_p^4}{4\xi^2} 
\left[1 + \exp\left(-\frac{2\phi}{\sqrt{6} m_p} \right) \right]^{-2} \;,
\]
where $\xi$ is a coupling constant between the Higgs field
and the scalar curvature.  As the authors' paper makes clear, 
it is not possible to have a rigorous discussion of
quantum corrections due to the nonrenormalizable nature of gravity.
The counting of speculations entering the Higgs inflation
model is:
\begin{itemize}
\item[-] Quantum gravity [F]
\item[-] Modification to gravity beyond GR [F]
\item[-] Sub-Hubble mass inflaton field [F]
\item[-] Symmetry not of type in SM - transform from the Jordan to Einstein frame [T]
\item[-] Symmetry included in the model [T]
\item[-] Model building beyond the Standard Model [T]
\end{itemize}
This gives a speculation count for Fundamental/Technical
properties of 3/3.


\medskip

\noindent
{\bf Warm little inflaton model
\cite{Bastero-Gil:2016qru,Bastero-Gil:2019gao}:} This 
model has two complex Higgs fields with identical
$U(1)$ charges. The fields have nonzero vacuum expectation values
and the phases of both fields then yield two
Nambu-Goldstone bosons.  The relative phase of the two fields
yields a singlet which is the inflaton.  The Higgs fields are coupled
to left-handed fermions with $U(1)$ charge and right-handed
counterparts that are gauge singlets.  There is an interchange
symmetry between the two bosons and two fermions
and they have identical couplings.  There is an additional
chiral fermion and singlet bosonic field to couple
with the fermions for the particle creation decay width.
The interaction Lagrangian for this model is 
given in Eq. (\ref{WLI_lagrangian}).
The inflaton potential for this model is simply monomials,
\[
V_{\rm warm \ little}(\phi) = \frac{\lambda}{4!} \phi^4  \ \ {\rm or} \ \ \frac{1}{2} m_{\phi}^2 \phi^2 \;.
\]
The counting of speculations entering the warm little inflaton inflation
model in the strong dissipative regime is:
\begin{itemize}
\item[-] Effective field theory methods with cutoff below $m_p$ [T]
\item[-] Symmetries included in the model - two Nambu-Goldstone bosons, two
$U(1)$, two interchange (6) [T]
\item[-] Extra fields beyond the SM not attributed to any symmetries (2) [T]
\item[-] Model building beyond the Standard Model [T]
\end{itemize}
This gives a speculation count for Fundamental/Technical
properties of 0/10.  This is the only model studied here with
no speculation counts in the fundamental category.
Note that in the weak dissipative regime the model would
have at least one and up to two fundamental 
counts for sub-Hubble mass and quantum
gravity, the latter because in cases there can be super-Planckian field
excursion of the inflaton field.  The speculation count highlights
the importance of the strong dissipative regime of warm inflation.
Nevertheless even in the weak dissipative regime there are less
speculation counts in the fundamental category than all
the other models examined here.


\bigskip

For the technical
speculation points in each model, one can now look
closer at the given model and see how well it can be parametrically
constrained. The speculation count is only assessing
the physical attributes of the model.  One would need
to look into the details of the inflation calculation in
the given model to see how constrained it is for inflation.
Alongside that one can also explore if other related cosmological
phenomenon such as baryongensis, dark matter, dark energy,
and cosmic magnetic fields can be explained by the model
or extensions of it. From this one can assess the full
parametric constraints on the model. Overall from
the models tested here, only the warm little inflaton
model has no fundamental speculation points and so is
in the best position for comparison to cosmological data.
Of course this model still should be further scrutinized but also
further model building can be explored to explain other cosmological
phenomenon.  Building inflation models involving
ideas from higher theories like quantum gravity etc...
can be interesting, but it is extremely challenging to
build an inflation model based on just the standard QFT that
we presently understand.  The warm little inflaton has
achieved this, so is a model of considerable interest.

The speculation count provides a measure beyond comparing to
observational data to assess not just inflation models but
cosmological models at large. Amongst the models that
agree well with the data, the speculation count
provides a semi-quantitative measure to disentangle the level
of assumption going into each of them.   
If the fundamental count is low and so is the technical
count, then there is a better chance comparing
that model to data can provide some insight into
the unknown fundamental attributes. However if
the technical count is also high, then such a model
can provide little such insight.
If the fundamental count is high, then irrespective
of the technical count, comparison to data will
provide very little insight into the unknown
fundamental attributes.  If there is no fundamental
count, then the model is in position to consider how well
the data constrains the parameters of the model and
whether the model can be used to provide other cosmological
predictions.

With the proliferation of inflation models, with so many
agreeing well enough with the limited cosmological early universe
data that it may as well be an infinite number,
more is needed to help separate out models in terms of
their qualities.
Theoretical cosmology
is different from most other fields in science in that
it requires a greater level of speculation to make any progress.
A speculation count is necessary
in theoretical cosmology, to keep cognizant of the levels
of assumption entering into model building, thus helping
to keep the field on an even scientific keel.
This is the needed `doubt' I believe Landau's statement in jest
was trying to convey that cosmologists lack.

This count should not be misunderstood as
trying to make some simple-minded separation
of models as good versus bad.
From a different way of thinking, purely in a mathematical physics
context, D-brane inflation is one of the most
elegant models. However is it a physically relevant model
or even close to that,
based on what we currently know about Nature?
In theoretical physics there is a notion that fundamental physical
theories should be mathematically elegant, but this notion must
be constrained first and foremost
by the theory showing adequate connection
with the physical world and being mathematically consistent.
As detailed earlier in this Section, cosmology experiments intrinsically
can provide very limited information about the early
universe.  For any theories with Fundamental attributes
from the Table, that cosmological information 
on its own is not adequate to
confirm the theory. It requires other independent empirical
information, alongside of course the theory being consistent.
Overstating the empirical relevance of string based models or
those based on other higher theories also does no justice
to the core theoretical work being done in those fields.
It just furthers the image of these subject areas
as out of touch with physics, since they are unable to distinguish between
pertinent empirical models versus those mainly
for theoretical interest.
Counting the number of physical attributes in a model that
are speculative, helps to assess which models are
most relevant for testing against experimental cosmological
data.  The count provides
a semi-quantitative measure on the degree of belief
of a model given the present theoretical understanding.
The higher a model's speculation count,
especially in the fundamental category and to
a lesser extent in the technical category,
the less useful is
the limited data able to select it out.  
This speculation count is time dependent.
In the past 50 years,
our confidence in the Standard Model has been
fully confirmed and there is greater cosmological
data, but amongst the fundamental
attributes in the Table, not much has changed. 
Thus the timescale of significant change
for the speculation count
is around or larger than a career
lifetime.

The speculation count helps to avoid a lot of arguments and
debates on nowhere issues, by quantifying the degree of theoretical
uncertainty about a given model, and it can be obtained by
a simple counting of assumptions entering in the model.
This little input can help us view cosmological
models a bit more objectively.
The swampland development from a couple years ago is a good
example where some objectivity is needed.
Within theoretical cosmology
that work caused considerable controversy, perhaps even a division
in the field.  
However it has to be appreciated that our understanding
of physics around the Planck scale is evolving to
the extent that at present there isn't even any experimental information,
which typically is essential for any development in
theoretical physics, at these extremely high energies.
One should expect, not be surprised, at further such
developments like swampland in the future, since that is an obvious
consequence of string
theory and more generally quantum gravity being unsolved.
Likewise the validity of models based on these ideas may also
change with time.

In any event,
superplanckian field excursions, which are not allowed
by the swampland conditions, still imply, irrespective
of the swampland conditions, that the model will
be susceptible to higher dimensional operator corrections
arising from quantum gravity. These corrections are unknown
and in such models one is forced to make
an assumption that they don't affect the model.
This is an uncontrolled assumptions as it relates
to quantum gravity,
and once such an assumption enters
model building, one may as well assume
quantum gravity can sort everything out, never mind the model.
Similar problems arise for models with
an inflaton mass less than the Hubble scale.
Thus, there is little point in venting one's ire on
the swampland conditions, which has only magnified
already known problems.
String theorists are just doing their job in
understanding a very difficult problem, and their views
probably also will further evolve over time.  It is not a simple-minded
question of right or wrong in regards issues at and above
the Planck scale.  It is about how much clarity
there is in the matter, and for the time being there
is very little, which thus also applies to any models at this scale.
This makes a speculation count useful in forcing
one to accept from the onset that
models with high counts, no matter how
insightful, have greater chance to eventually be found
theoretically incomplete or wrong.
At the other end, models with low speculation count
have less theoretical baggage, so more relevant to be tested
against experiment.  The ideal case is a speculation count of zero,
so below the $10 \ {\rm TeV}$ regime,
where the Standard Model explains everything.
We know for cosmology we have to go to a higher speculation
count, but
if inflation really is a viable theoretical
idea, then in order to feel certain it is connected to the physical
world, we need to find a model with low speculation count.
We can not call some possible future tensor mode discovery in the CMB
as inflation, if we don't even have a working, consistent,
physically relatable,
model of what this inflation is.

\section{Discussion}
\label{diss}

For the many researchers who have and continue to study warm inflation,
they work with confidence that data and theory indicate it
is headed in the right direction.
Today warm inflation is a strong contender in developing into a theory
of the early universe.  The strength of the underlying theoretical foundations
of warm inflation provide at least some confidence that
a tensor mode will eventually be detected.
In this review I discussed primarily the
early work done by me and the work I did with collaborators
in developing warm inflation.  Many researchers have
made valuable contributions in developing warm inflation,
including building interesting models 
\cite{Maia:1999yt,Taylor:2000ze,Chimento:2002us,Herrera:2006ck,Mimoso:2005bv,delCampo:2008vr,Cid:2007fk,delCampo:2007cia,Zhang:2009ge,Henriques:2009hq,Herrera:2010yg,Zhang:2013yr,Setare:2012fg,Sharif:2013zwa,Setare:2014qea,Sharif:2015vda,Goodarzi:2016iht,Motaharfar:2016dqt,Levy:2016jfh,Peng:2016yvb,Herrera:2017qux,Motaharfar:2017dxh,Mohammadi:2020vgs,Zhang:2020sbk,Reyimuaji:2020bkm,Motaharfar:2021egj,Samart:2021eph,Mohammadi:2021gvf,AlHallak:2021hwb,Zhang:2021zol,DAgostino:2021vvv,Bose:2022wla,Payaka:2022jtb,Kanno:2022flo,Deb:2022qxb},
first principles model building 
\cite{LopezNacir:2011kk,Hall:2004zr,Jeannerot:2006jj,Mohanty:2008ab,BuenoSanchez:2008nc,Cai:2010wt,Matsuda:2012kc,Mishra:2011vh,Visinelli:2011jy,Bastero-Gil:2013owa,Zhang:2014dja,Bastero-Gil:2014oga,Bastero-Gil:2015nja,Batell:2015fma,Cheng:2015oqa,Notari:2016npn,Ferreira:2017lnd,Rosa:2018iff,Sa:2020fvn,Laine:2021ego,Piccinelli:2021vbq,DeRocco:2021rzv,Agrawal:2022yvu,Kitazawa:2022gzk,Klose:2022rxh},
as well as models 
that follow the basic ideas of warm inflation
of particle production and associated
dissipative effects during inflation
\cite{Kofman:2004yc,Green:2009ds,Anber:2009ua,Barnaby:2009dd,Pearce:2017bdc,Almeida:2020kaq}.
Work on warm inflation has been done in
understanding the underlying quantum field theory dynamics
\cite{Berera:1998gx,Berera:2001gs,Lawrie:2002wm,Lawrie:2002zd,Lawrie:2004hs,Berera:2004kc,Moss:2008lkw,Bastero-Gil:2010dgy,Moss:2006gt,Aarts:2007ye,Bastero-Gil:2012akf}, 
examining density perturbations 
\cite{Hall:2003zp,Graham:2009bf,DeOliveira:2002wk,Lee:2003ed,Bastero-Gil:2011rva,Bastero-Gil:2014jsa,Visinelli:2014qla,Bastero-Gil:2019rsp,Qiu:2021ytc,Das:2022ubr},
studying non-gaussianity
\cite{Gupta:2002kn,Bartolo:2004if,Gupta:2005nh,Moss:2007cv,Moss:2011qc,Bastero-Gil:2014raa,Zhang:2014kwa,Zhang:2015zta,Mirbabayi:2022cbt},
and testing models to data 
\cite{Benetti:2016jhf,Hall:2004ab,Moss:2007qd,Panotopoulos:2015qwa,Visinelli:2016rhn,Jawad:2017gwa,Bastero-Gil:2017wwl,Gron:2018rtj,Berera:2018tfc,Gomes:2018uhv,Arya:2018sgw,Sheikhahmadi:2019gzs,Gron:2022lqj,Santos:2022exm,Montefalcone:2022owy,AlHallak:2022haa,Montefalcone:2022jfw}. 
There is work showing how warm inflation
can realize cosmic magnetic fields 
\cite{Berera:1998hv,Piccinelli:2013jua,Bastero-Gil:2015dag},
baryogenesis
\cite{Brandenberger:2003kc,Bastero-Gil:2011clw,Basak:2021cgk}, 
primordial blackholes 
\cite{Arya:2019wck,Bastero-Gil:2021fac,Correa:2022ngq,Ballesteros:2022hjk,Bhattacharya:2023ztw},
and address the gravitino problem
\cite{Taylor:2000jw,BuenoSanchez:2010ygd,Bartrum:2012tg}.
Studies have examined various aspects of
the warm inflation scenario and dynamics
\cite{Bastero-Gil:2017yzb,Bedroya:2019tba,deOliveira:1997jt,Bellini:1998ki,Billyard:2000bh,Donoghue:2000fk,Maia:2001zu,Berera:2000xz,Koh:2007rx,Battefeld:2008py,Bastero-Gil:2012vuu,Gim:2016uvv,Bastero-Gil:2016mrl,Sayar:2017pam,Graef:2018ulg,Harko:2020cev,Trivedi:2020ljd,Das:2020lut,Arya:2022xzc,Bertolami:2022knd,Gashti:2022pvu},
including for other cosmological problems 
\cite{Dymnikova:2000gnk,Alexander:2001dr,Lieu:2011rj,Bartrum:2014fla,Papageorgiou:2022prc}, with
application also of the warm inflation ideas of dissipation
during vacuum energy driven expansion to dark energy
\cite{Dimopoulos:2019gpz,Rosa:2019jci,Lima:2019yyv,DallAgata:2019yrr,Papageorgiou:2020nex}
with a possible resolution to
the Hubble tension \cite{Berghaus:2019cls}.
There are various reviews of warm inflation
\cite{Berera:2008ar,Bastero-Gil:2009sdq,Berera:2006xq,Gron:2016nxb,Rangarajan:2018tte}.  

Whether warm inflation, or inflation more generally, is the correct
idea about the early universe will be decided by empirical data.
Irrespective, warm inflation has introduced and substantiated
two concepts that are broadly useful to early universe
cosmology beyond just inflation and are here to stay.
One, that particle production from quantum field
interactions is possible at this early stage.
Aside from filling the universe with particles, the
other lesson to take from our work is the backreaction effects
of this process on the source are equally important in having
dynamical consequences for this early phase of the universe.
Two, that the initial primordial fluctuations could be of classical
rather than quantum origin, thus changing a way of thinking that
had dominated for decades, even before the inflation idea.
The warm inflation story is very much a part of the early universe cosmology
story, and those who have ignored it,
or the many other facets of this problem,
have lost the plot.

\vspace{-0.7cm}

\section*{Acknowledgements}

\vspace{-0.4cm}

I was partially funded by STFC.
For the purpose of open access, the author has applied a Creative Commons
Attribution (CC BY) licence to any Author Accepted Manuscript version
arising from this submission.

\end{document}